\documentclass{svjour3}
\usepackage{epsfig,amsmath,amsfonts,bm}
\usepackage{amssymb}
\usepackage[left=2.5cm,right=2.5cm,bottom=3.5cm,top=3.5cm]{geometry}
\usepackage{cuted}
\usepackage{standalone}
\usepackage{listings}
\usepackage{courier}

\usepackage[margin=10pt,font=footnotesize,labelfont=bf,labelsep=endash]{caption}
\let\OLDthebibliography\thebibliography
\renewcommand\thebibliography[1]{
  \OLDthebibliography{#1}
  \setlength{\parskip}{0pt}
  \setlength{\itemsep}{0pt plus 0.3ex}
}
\usepackage{subfiles}
\usepackage{subcaption}
\usepackage[linesnumbered,ruled,vlined]{algorithm2e}
\usepackage{hyperref}
\hypersetup{
    colorlinks=true,
    linkcolor=black,
    filecolor=magenta,
    urlcolor=cyan!70!black,
    bookmarks=true,
    citecolor=cyan!70!black
}
\usepackage{pgfplots}
\usepackage{filecontents}
\usepackage{tikz}
\usepackage{pgf}
\usetikzlibrary{patterns}
\usetikzlibrary{calc}
\usetikzlibrary{decorations.pathmorphing,decorations.markings}
\usepackage{nomencl}
\makenomenclature

\usepackage{etoolbox}
\renewcommand\nomgroup[1]{%
  \item[\bfseries
  \ifstrequal{#1}{K}{Equation of Motion Tensors}{%
  \ifstrequal{#1}{V}{Eigenvectors}{%
  \ifstrequal{#1}{C}{Correction vectors}{%
  \ifstrequal{#1}{R}{Reconstruction vectors}{}}}}%
]
}
\title{Direct computation of nonlinear mapping via normal form for reduced-order models of finite element nonlinear structures}
\titlerunning{Direct normal form for reduced-order models of finite element nonlinear structures}

\author{Alessandra Vizzaccaro
 \and
 Yichang Shen
 \and 
 Lo{\"i}c Salles
 \and
 Ji\v{r}\'i Blaho\v{s}
 \and Cyril Touz{\'e}}
\institute{A. Vizzaccaro \at
	Imperial College London, Exhibition Road, SW7 2AZ London, UK\\
	\email{a.vizzaccaro17@imperial.ac.uk}\\
	\and
	Y. Shen 	\at
	IMSIA, ENSTA Paris, CNRS, EDF, CEA, Institut Polytechnique de Paris, 828 Boulevard des Mar{\'e}chaux, 91762 Palaiseau Cedex, France
	\and
	L. Salles	\at
	Imperial College London, Exhibition Road, SW7 2AZ London, UK \\
	\and
	J. Blaho\v{s}\at
	Imperial College London, Exhibition Road, SW7 2AZ London, UK \\
	\and	
	C. Touz\'e	\at
	IMSIA, ENSTA Paris, CNRS, EDF, CEA, Institut Polytechnique de Paris, 828 Boulevard des Mar{\'e}chaux, 91762 Palaiseau Cedex, France
}

\newcommand{\modal}[1]{{\color{red!0!black}{#1}}}
\newcommand{\phys}[1]{{\color{blue!0!black} {#1}}}
\newcommand{\normal}[1]{{\color{green!0!black} {#1}}}

\newcommand{\x}{\modal{\mathbf{x}}}
\newcommand{\y}{\modal{\mathbf{y}}}
\newcommand{\X}{\phys{\mathbf{X}}}
\newcommand{\Y}{\phys{\mathbf{Y}}}
\newcommand{\R}{\normal{R}}
\renewcommand{\S}{\normal{S}}
\newcommand{\M}{\phys{\mathbf{M}}}
\newcommand{\K}{\phys{\mathbf{K}}}
\newcommand{\C}{\phys{\mathbf{C}}}
\newcommand{\zervec}{\phys{\mathbf{0}}}
\newcommand{\Og}{\phys{\mathbf{O}^2}}
\newcommand{\Ot}{\modal{\bm{\Omega}^2}}
\newcommand{\G}{\phys{\mathbf{G}}}
\newcommand{\g}{\modal{\mathbf{g}}}
\renewcommand{\H}{\phys{\mathbf{H}}}
\newcommand{\h}{\modal{\mathbf{h}}}
\newcommand{\e}{\modal{\mathbf{e}}}
\newcommand{\V}{\phys{\mathbf{V}}}
\newcommand{\I}{\modal{\mathbf{I}}}
\newcommand{\phit}{\phys{\bm{\phi}}}
\newcommand{\at}{\modal{\mathbf{a}}}
\newcommand{\bt}{\modal{\mathbf{b}}}
\newcommand{\ct}{\modal{\bm{\gamma}}}
\newcommand{\rt}{\modal{\mathbf{r}}}
\newcommand{\ut}{\modal{\mathbf{u}}}
\newcommand{\mt}{\modal{\bm{\mu}}}
\newcommand{\nt}{\modal{\bm{\nu}}}
\newcommand{\ag}{\phys{\bm{\bar{a}}}}
\newcommand{\bg}{\phys{\bm{\bar{b}}}}
\newcommand{\cg}{\phys{\bm{\bar{\gamma}}}}
\newcommand{\agd}{\phys{\bm{\bar{\alpha}}}}
\newcommand{\bgd}{\phys{\bm{\bar{\beta}}}}
\newcommand{\cgd}{\phys{\bm{\bar{c}}}}
\newcommand{\rg}{\phys{\bm{\bar{r}}}}
\newcommand{\ug}{\phys{\bm{\bar{u}}}}
\newcommand{\mg}{\phys{\bm{\bar{\mu}}}}
\renewcommand{\ng}{\phys{\bm{\bar{\nu}}}}
\newcommand{\EigVec}{\mathbf{\Phi}}
\newcommand{\EigVal}{\bm{\omega}}

\newcommand{\At}{\modal{\mathbf{A}}}

\newcommand{\Bt}{\modal{\mathbf{B}}}
\newcommand{\Ag}{\mathbf{\bar{A}}}

\newcommand{\Bg}{\mathbf{\bar{B}}}
\newcommand{\Z}{\phys{\mathbf{\bar{Z}}}}
\newcommand{\z}{\modal{\mathbf{Z}}}
\newcommand{\Zs}{\phys{\mathbf{\bar{Z}s}}}
\newcommand{\Zd}{\phys{\mathbf{\bar{Z}d}}}
\newcommand{\Zss}{\phys{\mathbf{\bar{Z}ss}}}
\newcommand{\Zdd}{\phys{\mathbf{\bar{Z}dd}}}
\newcommand{\Za}{\phys{\mathbf{\bar{Z0}}}}
\newcommand{\Zi}{\phys{\mathbf{\bar{Z1}}}}
\newcommand{\Zj}{\phys{\mathbf{\bar{Z2}}}}
\newcommand{\Zk}{\phys{\mathbf{\bar{Z3}}}}
\newcommand{\Zat}{\phys{\mathbf{{Z0}}}}
\newcommand{\Zit}{\phys{\mathbf{{Z1}}}}
\newcommand{\Zjt}{\phys{\mathbf{{Z2}}}}
\newcommand{\Zkt}{\phys{\mathbf{{Z3}}}}
\newcommand{\Pa}{\phys{\mathbf{\bar{P0}}}}
\renewcommand{\Pi}{\phys{\mathbf{\bar{P1}}}}
\newcommand{\Pj}{\phys{\mathbf{\bar{P2}}}}
\newcommand{\Pk}{\phys{\mathbf{\bar{P3}}}}
\newcommand{\Pol}[3]{\mathbf{\modal{#1}}_{\textcolor{black}{#2}}^\modal{(#3)}}
\newcommand{\pol}[3]{#1_{#2}^{(#3)}}
\begin{document}
\maketitle
\begin{abstract}
The direct computation of the third-order normal form for a geometrically nonlinear structure discretised with the finite element (FE) method, is detailed. The procedure allows to define a nonlinear mapping in order to derive accurate reduced-order models (ROM) relying on invariant manifold theory. 
The proposed reduction strategy is direct and simulation free, in the sense that it allows to pass from physical coordinates (FE nodes) to normal coordinates, describing the dynamics in an invariant-based span of the phase space. The number of master modes for the ROM is not a priori limited since a complete change of coordinate is proposed. The underlying theory ensures the quality of the predictions thanks to the  invariance property of the reduced subspace, together with their curvatures in phase space that accounts for the nonresonant nonlinear couplings. The method is applied to a beam discretised with 3D elements and shows its ability in recovering internal resonance  at high energy. Then a fan blade model is investigated and the correct prediction given by the ROMs are assessed and discussed. A method is proposed to approximate an aggregate value for the damping, that takes into account the damping coefficients of all the slave modes,  and also using the Rayleigh damping model as input. 
Frequency-response curves for the beam and the blades are then exhibited, showing the accuracy of the proposed method.
\end{abstract}


\section{Introduction}

Model order reduction methods for geometrically nonlinear structures is an active field of research and numerous techniques have been proposed in the past~\cite{ShawPierre91,Steindl01,mignolet13,Roberts2014,Haller2016}. Among them, nonlinear mappings and reduction to invariant manifolds attracts a special attention since they are able by nature  to provide better results than any linear method~\cite{ShawPierre93,TOUZE:JFS:2007,PONSIOEN2018}.  Indeed, the invariance property is key to ensure that trajectories from the ROM also exist for the full model, and the curvature of the reduction spaces allows the use of less coordinates to describe the dynamics than any linear decomposition.

Nonlinear normal modes (NNMs) offers a framework for such reduction methods. Since their first definition by Rosenberg in the 1960s~\cite{Rosenberg62}, they have witnessed numerous developments and definitions: family of periodic orbits~\cite{VakakisNNM,KerschenNNM09}, invariant manifold tangent at the origin to the linear eigenspaces~\cite{ShawPierre93,ShawPierre94}, as well as numerous computational methods, from asymptotic developments~\cite{ShawPierre93,nayfehnayfeh94,regacarbo00,touze03-NNM,Wagg2019} to numerical solution~\cite{Slater96,PesheckJSV,NORELAND2009,BLANC2013,RENSON2016} including continuation of periodic orbits~\cite{Lewandowki97b,ARQUIER2006,COCHELIN2009,PeetersNNM09}. Recently, Haller and coworkers revisited the mathematical definition of NNMs in order to settle down a unified framework~\cite{Haller2016}. They defined a spectral submanifold (SSM) as the smoothest member of an invariant manifold family, tangent to a linear eigenstructure. In the context of conservative systems, this definition reduces to the Lyapunov subcenter manifolds that are filled with periodic orbits, thus allowing unification of the different definitions given in the past, from Rosenberg to Shaw and Pierre approach.

Based on the introduction of invariant manifolds, NNMs have been used in order to derive reduced-order models for nonlinear dynamical systems and geometrically nonlinear structures, see {\em e.g.}~\cite{PesheckJSV,ApiwaFEMNL03}. A real approach to normal form has also been derived in order to make clear the connection with the invariant manifold approach, and generalise the derivation of reduced-order models thanks to a complete nonlinear change of coordinates~\cite{touze03-NNM,TOUZE:JSV:2006,TouzeCISM}. Successful  applications to shells have been reported with this method~\cite{TOUZE:CMAME:2008,touze-shelltypeNL}. SSMs have been developed with efficient parametrisation allowing to deal with large amplitude and produce efficient predictive models~\cite{PONSIOEN2018,BreunungHaller18}.

Application of these methods to finite-element (FE) structures with geometric distributed nonlinearity, remains however scarce~\cite{ApiwaFEMNL03,Touze:compmech:2014,givois2019}. On the other hand, other methods have been proposed in the FE community in order to tackle geometric nonlinearity, with a particular emphasis to non-intrusive or indirect methods~\cite{mignolet13}, that can be derived from any commercial FE software by using only standard operations, and thus without the need to implement new calculations at the elementary level. The STEP (Stiffness Evaluation Procedure) allows for a non-intrusive computations of the modal nonlinear coupling coefficients~\cite{muravyov,Perez2014},  but as such is not a reduction method since it allows only projecting the equations of motion onto the linear modes basis. Improvements of the method have been proposed, see {\em e.g.} the dual modes proposed in~\cite{KIM2013}, the combination of STEP with a POD (Proper Orthogonal Decomposition)~\cite{BalmasedaPODSTEP}, or a modified STEP where displacements are imposed on selected degrees of freedom only~\cite{KimCantilever,Vizza3d}. 
Implicit condensation and expansion (ICE) have been proposed in order to eliminate the high-frequency components~\cite{Hollkamp2008,kuether2015,FRANGI2019} thanks to a two-step procedure where a stress manifold is fitted after a series of static loading computations. Finally, modal derivatives, first introduced by Idelsohn and Cardona~\cite{IDELSOHN1985} have also been used with this objective of finding out a complementary projection subspace with faster convergence than linear modes~\cite{Weeger2016,WuTisoMD}. Recent extension proposes to use modal derivatives in a nonlinear mapping perspective to compute a quadratic manifold as reduction subspace~\cite{Jain2017,Rutzmoser}. 

Recent studies compare all these approaches mainly developed in the computational mechanics community to invariant and spectral submanifolds in order to better understand their mathematical foundations, advantages and drawbacks. Haller and Ponsioen derived general theorems showing that condensation and modal derivatives need a slow/fast assumption between master and slave coordinates in order to ensure convergence to a correct solution~\cite{HallerSF}. Elaborating on this idea, the quadratic manifold (QM) has been compared to the normal form approach, showing again how the QM tends to invariant manifold only when slow/fast assumption holds~\cite{Vizzaccaro:NNMvsMD}. A numerical criterion was established, stating that this slow/fast separation could be claimed as soon as the eigenfrequencies of the slave coordinates are 4 times larger than those of the master. A particular emphasis was also placed on both cases of static (SMD) and full modal derivatives (MD), showing how quadratic nonlinearities can be  treated incorrectly with the SMD approach, even if the slow/fast assumption is fulfilled. The ICE method has also been compared to invariant manifolds in~\cite{YichangICE}, again underlining the requirement of slow/fast separation. Importantly, ICE and QM methods produce reduction subspaces that are not velocity-dependent and neither invariant, and these simplifications become an important drawback in the prediction given by the ROMs~\cite{Vizzaccaro:NNMvsMD,YichangICE}.

These results advocate better application of reduction methods based on invariant manifold for FE structures as well, that would take properly into account the specificity of FE discretisation and also open the doors to possible non-intrusive procedures that could inherit the intrinsic properties of reduced subspace ({\em e.g.} invariance, velocity dependence). Also, a main advantage of invariant-based techniques is that they are by nature {\em simulation free}, {\em i.e.} without the need of resorting to a priori computations or outputs from the full-order model, as is the case for example for the POD or the PGD (Proper Generalized Decomposition) methods~\cite{KryslPOD,AmabiliPOD1,Chinesta2011,MEYRAND2019}. Such an effort has been recently tackled for methods based on SSM~\cite{VERASZTO}. By using a single master coordinate, the authors gave general formula, up to the third-order, allowing for a direct computation of a ROM from the physical coordinates. 
The proposed method is however limited to a single master mode in the proposed version. Even though the extension to multiple modes is theoretically possible, following {\em e.g.}~\cite{PesheckMultiNNM}, the procedure needs to be revised and developed further for direct computation.

The objective of this contribution is to extend the results obtained thanks to the real normal form approach~\cite{touze03-NNM,TOUZE:JSV:2006} to FE structures. A main limitation of the method as it was presented in~\cite{touze03-NNM,TOUZE:JSV:2006} is to rely on the equations of motion written in the modal basis. Since this projection step is out of reach for a number of FE problems involving very large number of degrees of freedom, the method has been completely rewritten in order to compute the needed quantities directly from the FE unknowns. In that manner, a full nonlinear mapping is introduced, allowing one to compute directly the reduced dynamics in an invariant-based span of the phase space, up to the third-order. 

The paper is organised as follows. Section~\ref{sec:theory} details the method, from the theoretical foundations to practical considerations on implementation. More particularly, Section~\ref{sec:frame} recalls the general framework on geometric nonlinearities used in this contribution, while Section~\ref{sec:NFmodal} briefly recalls the nonlinear mapping defined from the normal form approach, using modal coordinates as starting point. Then section~\ref{sec:NFo2} describes the calculations needed to perform the direct computation of the second-order normal form and gives its associated reduced dynamics. It also offers a short comparison to the quadratic mapping proposed in the framework of modal derivatives in order to understand the gain brought by the normal form approach. Section~\ref{sec:NFo3} extends the method to the third-order. All these results being obtained in the case of no internal resonance between the eigenfrequencies, section~\ref{sec:IR} explains how the method needs to be adapted to account for possible internal resonance. Since most of the engineering application are for forced and damped systems, the addition of external force and damping is detailed in section~\ref{sec:dampforc}. For the damping, a simplified expression is provided, that allows to take into account the case of lightly damped systems, and is specified to the case of Rayleigh damping, which is the most common form of losses customary introduced in FE codes. Section~\ref{sec:algo} closes the theoretical part by giving more details on the implementation and comments on the non intrusiveness characteristics of the technique, thanks to an algorithmic presentation. Section~\ref{sec:results} shows simulation results obtained on different structures. The conservative backbone curves (frequency-amplitude relationships) for a clamped-clamped beam and a fan blade are shown, illustrating the versatility of the method in taking into account internal resonances between nonlinear frequencies by adding more master coordinates in the ROM. Finally, frequency-response functions (FRFs) including damping and harmonic forcing are shown for both cases, illustrating how the method can handle these additional terms and produce efficient and reliable ROMs.

%

\section{Direct computation of normal form}\label{sec:theory}

\subsection{Framework}\label{sec:frame}

In this contribution, the case of geometric nonlinearity, corresponding to thin structures experiencing large amplitude vibrations, is considered. It is also assumed that the structure has been discretised with the finite element (FE) method.  The time-dependent displacement vector $\X$ gathers all the degrees of freedom (dofs) of the model (displacements/rotations at each nodes) and is  $N$-dimensional. The equation of motion reads:
\begin{equation}
\M\ddot{\X}+\K \X+\G(\X,\X)+\H(\X,\X,\X) = \zervec,
\label{eq:eom_phys}
\end{equation}
where $\M$ is the mass matrix, and $\K$ the tangent stiffness matrix. Geometric nonlinearity involves a nonlinear strain/displacement relationship and a linear elastic behaviour law. In the framework of three-dimensional elasticity, it implies that only quadratic and cubic polynomial terms have to be taken into account~\cite{mignolet13,LazarusThomas2012,Touze:compmech:2014,givois2019}, they are expressed thanks to the terms $\G(\X,\X)$ and $\H(\X,\X,\X)$, using a functional notation for the quadratic and cubic terms with coefficients gathered in third-order tensor $\G$ and fourth-order tensor $\H$. The explicit indicial expression  reads:
\begin{align}
\G(\X,\X)&=\sum^N_{r=1}\sum^N_{s=1}\G_{rs}\phys{X}_r\phys{X}_s,\\
\H(\X,\X,\X)&=\sum^N_{r=1}\sum^N_{s=1}\sum^N_{t=1}\H_{rst}\phys{X}_r\phys{X}_s\phys{X}_t,
\label{eq:tensor_product_phys}
\end{align}
where $\G_{rs}$ the N-dimensional vector of coefficients $G^p_{rs}$, for $p=1,\, ...,\, N$. 
In the first part of the paper, the main results are given for a conservative system, without taking into account a damping term in the equation of motion. However, section \ref{sec:dampforc} will be devoted to damping and forcing, and an approximate method will be derived in order to show how light damping and forcing can also be taken into account in the reduced-order modeling strategy.

For the following development, the eigenproblem needs to be defined. Let $(\omega_i,\phit_i)$ be the couple eigenfrequency-eigenvector verifying the  problem:
\begin{equation}
(\K - \omega_i^2 \M)\phit_i = \zervec.
\label{linear_eigsys}
\end{equation}
Using normalisation with respect to mass, then one has the following equations fulfilled:
\begin{equation}
\V^T\M\V=\I,\qquad\mbox{and}\qquad\V^T\K\V=\Ot,
\label{eq:diagonalisation}
\end{equation}
with $\V$ the matrix of all eigenvectors, $\V = (\phit_1, ..., \phit_N)$, $\I$ the identity matrix, and $\Ot$ a diagonal matrix composed of the square of the eigenpulsations, $\Ot = \mbox{diag}(\omega_i^2)$. Since most of the presented results will use previous calculations of normal form from the modal basis as a starting point, the equations of motion in the modal coordinates are used in the rest of the text in order to draw out parallels between the methods. Using the linear change of coordinates $\X = \V \x$, with $\x$ the $N$-dimensional vector of modal displacements, the dynamics reads:
\begin{equation}
\ddot{\x}+\Ot\x+\g(\x,\x)+\h(\x,\x,\x) = \zervec,
\label{eq:eom_modal}
\end{equation}
where the third- and fourth-order tensors $\g$ and $\h$ expresses the nonlinear modal coupling coefficients. They are linked to their equivalent $\G$ and $\H$ in the physical basis via:
\begin{subequations}
\begin{align}
&\g_{ij}
=\V^T
\G(\phit_i, \phit_j),
\label{eq:g_ij}\\
&\h_{ijk}
=\V^T
\H(\phit_i, \phit_j, \phit_k).
\label{eq:h_ijk}
\end{align}
\end{subequations}
 In this contribution, the internal force vector is assumed to derive from a potential, and  symmetry relationships exist between the nonlinear quadratic and cubic coefficients. For the quadratic coefficients, one has in particular that $\G(\phit_i, \phit_j) = \G(\phit_j, \phit_i)$, from which follows $\g_{ij}=\g_{ji}$, as well as $g^i_{jk} = g^j_{ki} = g^k_{ij}$, while for the cubic terms: $h^i_{jkl} = h^i_{jlk} = h^i_{klj} = h^i_{lkj}$.

The main difficulties for deriving reduced-order models for geometric nonlinearity are linked, in a FE context, with the number of dofs $N$ which can be prohibitively large. The number of coefficients involved in $\G$ and $\H$ tensors scales respectively as $N^3$ and $N^4$, and due to the distributed nature of the nonlinearity, all the oscillators are nonlinearly coupled. On the other hand, the nonlinear dynamics  exhibited by such structures are often simple so that they can be captured by elementary systems of coupled nonlinear oscillators. However, in order to derive efficient ROMs, the projection subspace giving rise to such low-order dynamics has generally a complex, curved shape in the phase space. Obtaining such a projection in a straightforward and simulation-free context is the aim of a nonlinear mapping that could lead directly from the FE discretisation to variables describing the dynamics in a curved manifold. In the next sections, such nonlinear mappings will be given and detailed. They are based on the normal form approach previously developed from the modal coordinates, see~\cite{touze03-NNM,TOUZE:JSV:2006,TouzeCISM}, and briefly recalled in section~\ref{sec:NFmodal}. 

\subsection{Normal form from modal coordinates}\label{sec:NFmodal}

This section recalls the main results obtained in~\cite{touzeLMA,touze03-NNM} for a conservative system, which has then been extended in~\cite{TOUZE:JSV:2006} in order to deal with modal damping. In these contributions, the starting point is the equations of motion in modal coordinates, Eq.~\eqref{eq:eom_modal}. The method has then been applied to partial differential equations (PDE) of beams, plate and shell models, see {\em e.g.}~\cite{touze04-NNMCompStruct} and~\cite{TOUZE:CMAME:2008} respectively for beams and shells, \cite{TOUZE:JFS:2007} for a comparison with the POD method, \cite{touze-shelltypeNL} for the correct prediction of the hardening/softening behaviour of shallow spherical shells, and~\cite{TouzeCISM} for an overview.

The nonlinear mapping is defined up to the third order and is based on an asymptotic expansion that can be pushed further if needed. The theory of normal form; Poincar{\'e} and Poincar{\'e}-Dulac's theorems, lay the foundation for its derivation. It is also defined for both displacements and velocities. Previous derivations showed that the results are equivalent to the center manifold approach used by Shaw and Pierre~\cite{ShawPierre91}. Hence the nonlinear mapping allows one to express the dynamics in an invariant-based span of the phase space. This property is crucial in order to defined accurate ROMs. Let us denote as $\y=\dot{\x}$ the velocity vector in physical space, $R_i$ and $S_j=\dot{R_j}$ the {\em normal} coordinates (respectively displacement and velocity). Let us also assume that the ROM is composed of $n$ master modes. Even if it is generally assumed that $n\ll N$, the method is given for a free $n$ and can also be used with $n=N$, hence defining a complete change of coordinates between  modal and  normal coordinates. The nonlinear mapping reads:
\begin{subequations}
\begin{align}
&\x=
\sum^n_i \e_i \R_i + 
\sum^n_{i=1}\sum^n_{j=1}
\at_{ij}
\R_i\R_j+
\sum^n_{i=1}\sum^n_{j=1}
\bt_{ij}
\S_i\S_j+
\sum^n_{i=1}\sum^n_{j=1}\sum^n_{k=1}
\rt_{ijk}
\R_i\R_j\R_k+
\sum^n_{i=1}\sum^n_{j=1}\sum^n_{k=1}
\ut_{ijk}
\R_i\S_j\S_k,
\label{eq:nonlinear_change_x}
\\
&\y=
\sum^n_i \e_i \S_i + 
\sum^n_{i=1}\sum^n_{j=1}
\ct_{ij}
\R_i\S_j+
\sum^n_{i=1}\sum^n_{j=1}\sum^n_{k=1}
\mt_{ijk}
\S_i\S_j\S_k+
\sum^n_{i=1}\sum^n_{j=1}\sum^n_{k=1}
\nt_{ijk}
\S_i\R_j\R_k.
\label{eq:nonlinear_change_y}
\end{align}
\label{eq:nonlinear_change_modal_compact}
\end{subequations}
In these expressions, $\e_i$ represents the unit eigenvector of the modal basis, with length $N$, composed of zeros except for the $i$-th entry which is equal to 1. The full expressions of all the reconstruction vectors $\at$, $\bt$, $\ct$, $\rt$, $\ut$, $\mt$, and $\nt$ are respectively given in Appendix~\ref{app:atobara} and \ref{app:rtobarr}, recalling the detailed  expressions demonstrated in~\cite{touze03-NNM}.
The sole difference between these expression and those provided in~\cite{touze03-NNM}, lies in the treatment of the symmetric terms in the summations. In these expressions, full summations are used with lower index covering from 1 to $N$. The other choice, often selected in such case, is to use upper-diagonalised forms for the tensors with ordered summations ($s\geq r$ and $t\geq s$), using the fact that the usual product is commutative. This choice is not retained here since it has been found easier to handle the expressions with full summations.

The following important properties of the nonlinear mapping from Eqs.~\eqref{eq:nonlinear_change_modal_compact} are recalled: (i) the change of coordinate is identity-tangent in the sense that the first order is colinear to a given eigenmode. The correcting terms allows to take into account the curvature of the NNM (invariant manifold). (ii) the number of master modes $n$ can be selected freely and in the most exhaustive case one can have $n=N$. The main advantage of these formula is that there is no need to recompute all the quantities when adding a new master mode. This is in contrast with methods based either on invariant manifolds, see {\em e.g.}~\cite{Boivin95}, or the recently proposed method based on SSM~\cite{VERASZTO}. 

The reduced-order dynamics is written for the general case where no internal resonance exists between the eigenfrequencies of the system. When an internal resonance is present, some terms are vanishing in Eqs.~\eqref{eq:nonlinear_change_modal_compact}, leading to extra terms staying in the normal form of the system. The theory is detailed in~\cite{touze03-NNM,TOUZE:JSV:2006,TouzeCISM}, and further comments will be provided in the next sections. In any case the dynamics onto the third-order $2n$-dimensional invariant manifold corresponding to the $n$ selected master modes writes, in this general case without internal resonance, $\forall r=1...n$:
\begin{equation}
\begin{split}
&
\ddot{\R}_r+\omega^2_r\R_r+
(\modal{A}^r_{rrr}+\modal{h}^r_{rrr})\R_r^3+
(\modal{B}^r_{rrr})\R_r\dot{\R}_r^2\\
&
+\R_r
\sum^n_{j\neq r}
(\modal{A}^r_{jjr}+\modal{A}^r_{jrj}+\modal{A}^r_{rjj}+3\modal{h}^r_{rjj})\R_j^2
+\R_r
\sum^n_{j\neq r}
(\modal{B}^r_{rjj})\dot{\R}_j^2
+
\dot{\R}_r
\sum^n_{j\neq r}
(\modal{B}^r_{jjr}+\modal{B}^r_{jrj})\R_j\dot{\R}_j
=0.
\end{split}
\label{eq:ROM}
\end{equation}
As clearly emphasised in Eq.~\eqref{eq:ROM} where $R_r$ and $\dot{R}_r$ terms have been factorised, the reduced dynamics does not contain invariant-breaking terms anymore.  An important remark is also that  since no internal resonance  have been assumed between the eigenfrequencies, no quadratic terms are present in Eq.~\eqref{eq:ROM}, only cubic terms corresponding to trivial  resonance stay in the normal form. Finally, one can note  the presence of velocity-dependent terms in the reduced dynamics, reflecting the velocity-dependence of  the invariant manifolds. Even though displacement and velocities are used as independent variables in the process, the method can simply reduce to displacements only, one has just to replace  $S_r=\dot{R}_r$ everywhere to obtain full expressions depending on displacements only. In the same line, Eq.~\eqref{eq:nonlinear_change_y} allows for a better reconstruction of the velocities but is not mandatorily needed and can be deduced from~\eqref{eq:nonlinear_change_x}.

New fourth-order tensors $\At$ and $\Bt$ appear in these equations, their expressions from the modal basis can be found in~\cite{touze03-NNM} and are here recalled:
\begin{subequations}\label{eq:ABmodal}
\begin{align}
\At_{ijk}=\sum^N_{s=1} (\g_{is}+\g_{si})\modal{a}^s_{jk}=\sum^N_{s=1}2\,\g_{is}\modal{a}^s_{jk},\\
\Bt_{ijk}=\sum^N_{s=1} (\g_{is}+\g_{si})\modal{b}^s_{jk}=\sum^N_{s=1}2\,\g_{is}\modal{b}^s_{jk},
\end{align}
\end{subequations}
where the last simplification stems from the symmetry of the quadratic tensor $\g$. 

In a FE context, the main problem is that applicability of this reduction procedure needs as a first step the full equations in modal basis, Eq.~\eqref{eq:eom_modal}, with all the coefficients $\g$ and $\h$ known and computed. Even though theoretically feasible by resorting to the {\em Stiffness Evaluation Procedure} (STEP) as described in~\cite{muravyov}, this computation is generally out of reach for much of the FE structures since the associated computational cost is prohibitively large. Also, an {\em a priori} selection of some of the important coupled eigenmodes, which could indicate a way of solving that dimensionality issue, is clearly out of reach, as underlined for example for 3D elements in~\cite{Vizza3d}, where non-negligible couplings with very high frequency thickness modes have been exhibited. Hence there is a clear  need for a direct computation of the normal form from the FE discretisation, since it will allow one to take directly into account all the coupled modes without any a priori or assumptions to formulate, and will define a simulation-free method applicable for efficient reduced-order modeling with geometric nonlinearity.

\subsection{Second-order direct normal form}\label{sec:NFo2}

In order to introduce progressively the cubic nonlinear mapping, we begin with a first step where only the second-order terms of the normal form are computed directly from the FE discretisation. This first step allows for a better understanding of the physical meaning of the involved coefficients, as well as some short comparisons with the quadratic mapping introduced from the modal derivatives in~\cite{Jain2017,Rutzmoser}. The reduced dynamics in this case will also be specified and the meanings of this assumption further detailed. In section~\ref{sec:results} where numerical results will be provided, this assumption will be again discussed and highlighted on given test cases.

The quadratic mapping, deduced from Eq.~\eqref{eq:nonlinear_change_modal_compact}, from the physical $\X$ coordinates of the FE model, reads:
\begin{subequations}\begin{align}
&\X=
\sum^n_{i=1} \phit_i \R_i + 
\sum^n_{i=1}\sum^n_{j=1}
\ag_{ij}
\R_i\R_j+
\sum^n_{i=1}\sum^n_{j=1}
\bg_{ij}
\S_i\S_j,
\label{eq:nonlinear_change_x_physO2}
\\
&\Y=
\sum^n_{i=1} \phit_i \S_i + 
\sum^n_{i=1}\sum^n_{j=1}
\cg_{ij}
\R_i\S_j.
\label{eq:nonlinear_change_y_phys02}
\end{align}
\label{eq:nonlinear_change_phys02}
\end{subequations}

In these expressions, $n$ stands for the number of master modes retained for building the ROM, and $\phit_i$ is the corresponding eigenvector. The {\em normal} coordinates $(\R_i,\S_j)$ have the same meaning as in Section~\ref{sec:NFmodal}, since they describe the dynamics in the same invariant-based span of the phase space. The expressions of the reconstruction vectors $\ag$, $\bg$ and $\cg$,  are the equivalent to those obtained in modal basis, the only difference being that the new ones can now be computed directly from the FE nodes. As already remarked in  section~\ref{sec:NFmodal}, the second equation~\eqref{eq:nonlinear_change_y_phys02} is not mandatorily needed for the reduction process. Indeed, \eqref{eq:nonlinear_change_y_phys02} can be deduced from \eqref{eq:nonlinear_change_x_physO2}, by adding the first-order assumption $\dot{\S}_p = -\omega^2_r \R_r$.

The detailed expressions of the second-order tensors $\ag$, $\bg$, $\cg$, are now given. They have been derived from the coefficients obtained in~\cite{touze03-NNM} where the starting point was the modal coordinates. Appendix~\ref{app:atobara} explains  how this operation is done with a particular emphasis on the $\ag$ tensor, also giving important computational details with regard to the direct computation in a FE context. Let us first define the two vectors $\Zs_{ij}$ and $\Zd_{ij}$ as:
\begin{subequations}\label{eq:ZsZd}
\begin{align}
\Zs_{ij}=((+\omega_i+\omega_j)^2\M-\K)^{-1} \G(\phit_i,\phit_j),\\
\Zd_{ij}=((-\omega_i+\omega_j)^2\M-\K)^{-1} \G(\phit_i,\phit_j),
\end{align}
\end{subequations}
which are needed to arrive at a compact expression for $\ag$, $\bg$ and $\cg$. These vectors encompass all the possible second-order internal resonance thanks to the appearance of the sum $\omega_i+\omega_j$ and difference $-\omega_i+\omega_j$ of two eigenfrequencies (thus giving the names of these $\Z$ vectors with $s$ (summation) and $d$ (difference) as subscripts). Indeed, the matrix to invert will become singular in case of existence of a second-order internal resonance. These terms come from the denominators of the modal normal form and vanish in case of internal resonance, see~\cite{touze03-NNM,TouzeCISM} and section~\ref{sec:IR} for more details. They have  been written  with these expressions because they are more convenient on the computational viewpoint: $\Zs_{ij}$ and $\Zd_{ij}$ can be computed easily  by solving a linear system, thus avoiding the matrix inversion. This will be further commented in  Section \ref{sec:algo} where specific algorithmic details will be highlighted.


The expressions for the three vectors $\ag$, $\bg$ and $\cg$  deduce from the two $\Zs_{ij}$ and $\Zd_{ij}$ only, underlining again the fact that the second equation on $\Y$ is not independent from the first one, so that in a simplified version of the method it can be neglected. They read:
\begin{subequations}\label{eq:abcphys}
\begin{align}
\ag_{ij}&=\dfrac{1}{2}(\Zd_{ij}+\Zs_{ij}), \label{eq:a_phys}\\
\bg_{ij}&=\dfrac{1}{2\omega_i\omega_j}(\Zd_{ij}-\Zs_{ij}), \label{eq:b_phys}\\
\cg_{ij}&=\dfrac{\omega_j-\omega_i}{\omega_j}\Zd_{ij}+\dfrac{\omega_j+\omega_i}{\omega_j}\Zs_{ij}. \label{eq:c_phys}
\end{align}
\end{subequations}

Vectors $\ag_{ij}$, $\bg_{ij}$ and $\cg_{ij}$ can  be seen as correction vectors needed to take into account the quadratic curvature of the invariant manifold. They thus share a common interpretation with the concepts of modal derivatives, as it has been introduced {\em e.g.} in~\cite{IDELSOHN1985,Weeger2016} . Indeed modal derivatives aimed at taking into account the nonlinear dependence of an eigenvector with respect to perturbations in other modal directions. It has been recognised  that this can be linked to the Hessian of the internal force vector and represents a manifold curvature in phase space~\cite{Jain2017,Rutzmoser,Vizzaccaro:NNMvsMD}. In order to make the connection with {\em static} modal derivative (SMD) more clear, let us assume that a single master coordinate, say $i$, is retained for building the ROM. Restricting to a single master mode motion, then Eqs.~\eqref{eq:ZsZd} simplifies to:
\begin{subequations}\label{eq:ZsZd_single}
\begin{align}
\Zs_{ij} & =((2\omega_i)^2\M-\K)^{-1} \G(\phit_i,\phit_i), \label{eq:ZsZd_singlea}\\
\Zd_{ij} & =-\K^{-1} \G(\phit_i,\phit_i). \label{eq:ZsZd_singleb}
\end{align}
\end{subequations}
One can observe that Eq.~\eqref{eq:ZsZd_singleb} is fully equivalent to the usual definition of static modal derivative $\bm{\theta}_{ii}$ one can found {\em e.g.} in~\cite{Weeger2016,Jain2017,Vizzaccaro:NNMvsMD}, to the multiplicative factor 2, {\em i.e.} $\Zd_{ii}=2\,\bm{\theta}_{ii}$. This means that in this simplified case, $\bm{\theta}_{ii}$ exactly recovers the correct direction in phase space that express the quadratic couplings.   However, this is not true anymore for cross-coupled SMD $\bm{\theta}_{ij}$, $i\neq j$, since SMD is not able to make appear the difference between eigenfrequencies, nor the sum.

A quite similar comment can also be addressed about $\Zs_{ii}$, with the additional assumption of a slow/fast separation between master and slave coordinates. Indeed, Eq.~\eqref{eq:ZsZd_singlea} makes appear implicitly, in the terms of the matrix to inverse, the differences $(2\omega_i)^2-\omega_s^2$ between the master coordinate~$i$ and all the slave modes~$s$. The slow/fast assumption requires that the slave modes are much more stiff than the master, implying $\forall s, \quad \omega_s\gg\omega_i$. Hence if this assumption is well fulfilled, then $\Zs_{ii}$ will also tend to the direction pointed by  $\Zd_{ii}=2\,\bm{\theta}_{ii}$. The quadratic manifold produced by SMD will then tend to the one provided by direct normal form, since $\ag$ will tend to twice the SMD and $\bg$ to zero. This simple comparison underlines the common point between quadratic manifold from SMD and the one proposed from direct normal form. In the general case, one understands that the formulas given by Eqs.~\eqref{eq:abcphys} are more general, and only degenerate to the formulas proposed in~\cite{Jain2017,Rutzmoser} when specific assumptions are met.
As a conclusion, one can state that $\ag$ and $\bg$ vectors are in fact the correct corrections that would have needed to be defined as the modal derivatives, since it gives the proper curvatures of the invariant manifolds (NNMs) which are defined as the continuation of the underlying eigenmodes, tangent to the linear subspaces at origin.  The interested reader is also referred to~\cite{Vizzaccaro:NNMvsMD} where a complete comparison between the two methods is provided.

%

In the case considered in this section where a second-order normal form is studied, it is also interesting to derive the reduced-order dynamics arising from this choice, which will express the dynamics onto second-order approximations of the invariant manifolds. The dynamics is nonetheless expressed up to cubic order and reads, $\forall \, r \, \in \; [1,n]$:
\begin{equation}
\ddot{\R}_r+\omega^2_r\R_r+
\sum^n_{i=1}\sum^n_{j=1}\sum^n_{k=1}
(A^r_{ijk} +   h^r_{ijk})  \R_i\R_j\R_k
+
B^r_{ijk}\R_i\dot{\R}_j\dot{\R}_k
=0.
\label{eq:ROM_SO}
\end{equation}

As compared to Eq.~\eqref{eq:ROM}, this reduced dynamics does not contain quadratic terms also, as a consequence of the fact that no quadratic terms are trivially resonant. As long as no second-order internal resonance exists, these terms can be cancelled out. However all the cubic terms are present, without any simplification. This is the main contrast with Eq.~\eqref{eq:ROM}, where all the non-trivial resonant monomial terms have been cancelled thanks to the third-order term in the nonlinear mapping. Consequently Eq.~\eqref{eq:ROM} contains much less cubic terms than~\eqref{eq:ROM_SO}, and in particular all the invariant coupling terms can be cancelled. This point will be further commented in sections~\ref{sec:NFo3} and~\ref{sec:results}. One can already note that reduced-order dynamics can be simulated with Eq.~\eqref{eq:ROM_SO}, meaning that: (i) the invariant manifolds are approximated to the second-order only, (ii) all the resonant monomial terms, including non-trivial ones, are present. This might appear interesting in certain cases where strong higher order internal resonance are present, especially when they involve resonance between the nonlinear frequencies, that one cannot easily foresee from a linear analysis. This point will be highlighted in section~\ref{sec:beam}.

In the particular case where a single master mode motion is selected, it is important to notice that Eqs.~\eqref{eq:ROM} and \eqref{eq:ROM_SO} reduce to the same, so that the reduced-order dynamics on a single invariant manifold is equivalent  with second and third-order normal form. Assuming that the master coordinate has label $r$, then $\forall k \neq r, \; \R_k=\S_k=0$, and the dynamics of $\R_r$ writes:
\begin{equation}
\ddot{\R}_r + \omega_r^2 \R_r + (A_{rrr}^r + h_{rrr}^r) \R_r^3 + B_{rrr}^r \R_r \dot{\R}_r^2 \; = \; 0 \; . 
\label{eq:dynsingledof}
\end{equation}
Indeed, in this simple case, the only cubic monomial terms staying in \eqref{eq:dynsingledof} are trivially resonant, and cannot be cancelled by the cubic order terms from the nonlinear mapping. This means that if one is interested to the reduction to a single invariant manifold, then there is no need to compute further the cubic terms from the normal form.

The last point to address is the direct computation of the $\At$ and $\Bt$, expressed in Eqs.~\eqref{eq:ABmodal}, and appearing in the reduced dynamics. Since their definitions make appear the $\g$  tensor of quadratic coefficients from the modal basis, a direct computation is needed to circumvent the problem of computing all of these  coefficients. For that purpose, let us define  $\Ag$ and $\Bg$ the equivalent tensors in physical coordinates of $\At$ and $\Bt$, which can be simply obtained by  premultiplying $\At$ and $\Bt$ by $\V^{-\text{T}}$. On the physical point of view, they are homogeneous to  forces and not to displacements. Indeed, the linear change of coordinates for displacements reads $\X=\V\x$, whereas for a force one has  $\modal{\bm{f}}=\V^{\text{T}}\phys{\bm{F}}$, where $\bm{f}$ is the modal force and $\bm{F}$ the nodal force. Since, by definition of the quadratic tensors in modal and physical coordinates, one has $\V^{-\text{T}} \g_{is} = \G (\phit_i, \phit_s)$, then:
\begin{subequations}
\begin{align}
\Ag_{ijk}=\sum^N_{s=1}2\, \G (\phit_i, \phit_s)\,  \modal{a}^s_{jk},\\
\Bg_{ijk}=\sum^N_{s=1}2\, \G (\phit_i, \phit_s)\, \modal{b}^s_{jk}.
\end{align}
\end{subequations}
To arrive at a finalised expression involving only terms from the physical basis,  one has to notice that   $\sum^N_{s=1}\phit_s\modal{a}^s_{jk} = \ag_{jk}$ (and similarly for $\bg$), by again using the simple projection rule. The coefficients $\modal{a}^s_{jk}$ and $\modal{b}^s_{jk}$ can then be brought into the brackets  of  $\G (\phit_i, \phit_s) $ leading to the final expressions:
\begin{subequations}\label{eq:ABphys}
\begin{align}
\Ag_{ijk}=2\, \G( \phit_i, \ag_{jk}),\\
\Bg_{ijk}=2\, \G( \phit_i, \bg_{jk}).
\end{align}
\end{subequations}
Eqs.~\eqref{eq:ABphys} show how $\Ag$ and $\Bg$ can be computed directly by simple manipulations of the FE code (more algorithmic details will be given in Section~\ref{sec:algo}). Once they have been obtained, their equivalent in modal basis simply read:
\begin{subequations}\label{eq:ABphysproj}
\begin{align}
\At_{ijk}=\EigVec^\text{T}\Ag_{ijk},\\
\Bt_{ijk}=\EigVec^\text{T}\Bg_{ijk},
\end{align}
\end{subequations}
where $\EigVec$ is the reduced matrix of the master eigenvectors, which should not be confused with $\V$:  $\EigVec$ contains only the $n$ master modes extracted from $\V$, whereas $\V$ contains all the $N$ eigenvectors.

To conclude this section, we have derived general second-order formula that give rise to a quadratic mapping based on the second-order normal form theory. The expressions have been compared to previous works using static modal derivatives and the difference between the two methods have been underlined. In particular, it has been shown how the proposed direct normal form (DNF) computation generalise the quadratic manifold SMD approach and allows to take properly into account the internal resonance, the curvature of the phase space and the velocity dependence. In the next section, we push the method further and give the detailed formulas needed to obtain the cubic terms of the nonlinear mapping.

\subsection{Third Order direct normal form}\label{sec:NFo3}
The computation of the third-order tensors is more difficult for two different reasons. First the expressions are longer and more tedious. Second and most importantly, at the cubic order, one has to distinguish between trivial and non-trivial internal resonance. The reader is referred to~\cite{touze03-NNM,TOUZE:JSV:2006,TouzeCISM} for detailed discussions, while sections~\ref{sec:TR} and \ref{sec:IR} recalls important definitions. Roughly speaking, if at second-order all the terms can be cancelled as long as no order-two internal resonance exists, this is not the case anymore at the third-order, otherwise this would mean that the normal form (the reduced-order dynamics) is linear. As explained {\em e.g.} in~\cite{TouzeCISM}, cubic monomial term $X_p^3$ on oscillator $p$ is trivially resonant, and on the physical point of view this is meaningful since it will bend the frequency-response curves so as to produce hardening or softening behaviour~\cite{touze03-NNM}. 

The nonlinear mapping up to order three reads:
\begin{subequations}\begin{align}
&\X=
\sum^n_{i=1} \phit_i \R_i + 
\sum^n_{i=1}\sum^n_{j=1}
\ag_{ij}
\R_i\R_j+
\sum^n_{i=1}\sum^n_{j=1}
\bg_{ij}
\S_i\S_j,
+
\sum^n_{i=1}\sum^n_{j=1}\sum^n_{k=1}
\rg_{ijk}
\R_i\R_j\R_k+
\sum^n_{i=1}\sum^n_{j=1}\sum^n_{k=1}
\ug_{ijk}
\R_i\S_j\S_k,
\label{eq:nonlinear_change_x_physO3}
\\
&\Y=
\sum^n_{i=1} \phit_i \S_i + 
\sum^n_{i=1}\sum^n_{j=1}
\cg_{ij}
\R_i\S_j
+
\sum^n_{i=1}\sum^n_{j=1}\sum^n_{k=1}
\mg_{ijk}
\S_i\S_j\S_k+
\sum^n_{i=1}\sum^n_{j=1}\sum^n_{k=1}
\ng_{ijk}
\S_i\R_j\R_k.
\label{eq:nonlinear_change_y_phys03}
\end{align}
\label{eq:nonlinear_change_phys03}
\end{subequations}

Even though the second equation on $\Y$ is not mandatorily needed in the reduction technique, it is given for the following reasons. First, the original method has been developed including this second equation. Even though it can be deduced from the first in order to have expressions involving only displacements, the needed approximation $\dot{\S}_r = -\omega_r^2 \R_r$ is just the leading-order and might encounter limitations when the third order is included. Second, giving Eq.~\eqref{eq:nonlinear_change_y_phys03} allows for direct reconstruction of the velocities, without deducing it from the first with an assumption, and  for a light added computational effort. Finally, full dependency on the second equation from the first could be lost in more complex cases, as {\em e.g.} including the damping (see~\cite{TOUZE:JSV:2006} for the expressions of the mapping   where all the monomials are involved) or other forces in the starting equation. All these points need further analytical investigations that are beyond the scope of the present study. For all these reasons, we keep the change of coordinates with both displacements and velocities. 

The expressions of the newly introduced coefficients $\rg_{ijk}$, $\ug_{ijk}$, $\mg_{ijk}$, $\ng_{ijk}$ are given in the next subsection. The presentation will first consider the simplest case of the terms corresponding to non-trivial internal resonance. These terms can always be  cancelled so that no specific treatment is needed as compared to the calculations presented in the previous section. Then the trivially resonant monomials are considered. These terms should not appear in the mapping so that the corresponding monomials will stay in the reduced dynamics (normal form). This specific treatment is detailed in section~\ref{sec:TR}.

The reduced dynamics after the cubic mapping is given by Eq.~\eqref{eq:ROM} in case of no internal resonance. It differs from \eqref{eq:ROM_SO} (reduced dynamics after second-order normal form) only by the fact that all the non resonant cubic terms have been cancelled. In some sense the order-three reduced dynamics   appears simpler after the cubic treatment, and contains less possible nonlinear interactions.
As a matter of fact, since the reduced dynamics is truncated at order three, the third-order treatment shown here is somehow incomplete. Indeed, non-resonant cubic monomial terms have been discarded, but the higher-order terms stemming from the nonlinear change of coordinates (quartic and quintic terms), have not been taken into account. We will show in section~\ref{sec:results} that this might have some important consequences when dealing with higher-order internal resonance. One must keep in mind that processing the cubic terms creates quartic terms that have not been considered in the present study, thus reducing the quality of the third-order reduced dynamics. A full treatment would require to go to higher orders, for example for a correct treatment of 5:1 resonance. Hence the third-order must be viewed as the needed first step toward this achievement, which is however beyond the scope of the present study.


\subsubsection{General treatment for non-trivially resonant monomials}

In this subsection, the non-resonant terms are first considered. They correspond to the cases where none of the possible combinations of the frequencies $\pm \omega_i \pm\omega_j \pm\omega_k$ is equal to another eigenfrequency $\omega_s$ of the system. The general expressions for the fourth-order tensors $\rg$, $\ug$, $\mg$, $\ng$ appearing in Eq.~\eqref{eq:nonlinear_change_phys03} have been obtained following a similar analysis as the one shown in Appendix~\ref{app:atobara} for the quadratic terms. This analysis is detailed in Appendix~\ref{app:rtobarr}. They are here expressed from four independent vectors $\Za_{ijk}$, $\Zi_{ijk}$, $\Zj_{ijk}$ and $\Zk_{ijk}$ given in Eqs.~\eqref{eq:ZaZiZjZk}, in a similar fashion as the quadratic tensors have been computed from $\Zs_{ij}$ and $\Zd_{ij}$ in the previous section.

To simplify the presentation, it is convenient to define the following vectors obtained as a combination of cubic forces:
\begin{subequations}\label{eq:PaPiPjPk}
\begin{align}
\Pa_{ijk}=
\Ag_{ijk}+\Ag_{jki}+\Ag_{kij}+3\H(\phit_i,\phit_j,\phit_k)
-\omega_j\omega_k\Bg_{ijk}-\omega_k\omega_i\Bg_{jki}-\omega_i\omega_j\Bg_{kij},
\\
\Pi_{ijk}=
\Ag_{ijk}+\Ag_{jki}+\Ag_{kij}+3\H(\phit_i,\phit_j,\phit_k)
-\omega_j\omega_k\Bg_{ijk}+\omega_k\omega_i\Bg_{jki}+\omega_i\omega_j\Bg_{kij},
\\
\Pj_{ijk}=
\Ag_{ijk}+\Ag_{jki}+\Ag_{kij}+3\H(\phit_i,\phit_j,\phit_k)
+\omega_j\omega_k\Bg_{ijk}-\omega_k\omega_i\Bg_{jki}+\omega_i\omega_j\Bg_{kij},
\\
\Pk_{ijk}=
\Ag_{ijk}+\Ag_{jki}+\Ag_{kij}+3\H(\phit_i,\phit_j,\phit_k)
+\omega_j\omega_k\Bg_{ijk}+\omega_k\omega_i\Bg_{jki}-\omega_i\omega_j\Bg_{kij}.
\end{align}
\end{subequations}
From them, one can obtain four independent cubic vectors of displacement that will be used to build the tensors:
\begin{subequations}
\begin{align}
\Za_{ijk}=((+\omega_i+\omega_j+\omega_k)^2{\M}-{\K})^{-1}
{\Pa}_{ijk},
\\
\Zi_{ijk}=((-\omega_i+\omega_j+\omega_k)^2{\M}-{\K})^{-1}
{\Pi}_{ijk},
\\
\Zj_{ijk}=((+\omega_i-\omega_j+\omega_k)^2{\M}-{\K})^{-1}
{\Pj}_{ijk},
\\
\Zk_{ijk}=((+\omega_i+\omega_j-\omega_k)^2{\M}-{\K})^{-1}
{\Pk}_{ijk}.
\end{align}\label{eq:ZaZiZjZk}
\end{subequations}
Note that in the computational algorithm, each of these vectors is obtained by solving a linear system of equations, and not by actually inverting the matrices, which is much more meaningful on the computational point of view. The matrices of these systems are  linear combinations of $\K$ and $\M$ and, similarly to the $\Zs$ and $\Zd$ vectors of the second order, cover all the possible combinations of $\omega_i$, $\omega_j$, and $\omega_k$. Four of them are needed in order to obtain all possible cases, corresponding to the splitting of the denominators appearing in the modal normal form approach. The case where one of these combinations is equal to another eigenfrequency $\omega_s$ of the system, thus making one of the matrices singular, will be specifically treated in Sec.~\ref{sec:TR}. Here we assumed that it is not the case, therefore the vectors $\Z$ can be computed.

Finally the expressions of the cubic tensors read:
\begin{subequations}
\begin{align}
&\rg_{ijk}=\dfrac{1}{12}\left(\Za_{ijk}+\Zi_{ijk}+\Zj_{ijk}+\Zk_{ijk} \right),\\
&\ug_{ijk}=\dfrac{1}{4\omega_j\omega_k}\left(-\Za_{ijk}-\Zi_{ijk}+\Zj_{ijk}+\Zk_{ijk} \right),\\
&\mg_{ijk}=\dfrac{1}{12\omega_i\omega_j\omega_k}\left(
-(+\omega_i+\omega_j+\omega_k)\Za_{ijk}
+(-\omega_i+\omega_j+\omega_k)\Zi_{ijk}
+(+\omega_i-\omega_j+\omega_k)\Zj_{ijk}
+(+\omega_i+\omega_j-\omega_k)\Zk_{ijk} \right),\\
&\ng_{ijk}=\dfrac{1}{4\omega_i}\left(
+(+\omega_i+\omega_j+\omega_k)\Za_{ijk}
-(-\omega_i+\omega_j+\omega_k)\Zi_{ijk}
+(+\omega_i-\omega_j+\omega_k)\Zj_{ijk}
+(+\omega_i+\omega_j-\omega_k)\Zk_{ijk} \right).
\end{align}\label{eq:third-order_tensors}
\end{subequations}

Before generalizing these computations to the case of trivial internal resonance, one can note that the $\Z$ vectors introduced in Eqs.\eqref{eq:ZaZiZjZk} (where $\Z$ is a shortcut notation for all $\Za_{ijk}$, $\Zi_{ijk}$, $\Zj_{ijk}$ and $\Zk_{ijk}$) have some symmetry relationships:
\begin{subequations}
\begin{align}
&\Za_{ijk} = \Za_{ikj} = \Za_{jki} = \Za_{jik} = \Za_{kij} = \Za_{kji},\\
&\Zi_{ijk} = \Zi_{ikj} = \Zk_{jki} = \Zj_{jik} = \Zj_{kij} = \Zk_{kji},\\
&\Zj_{ijk} = \Zk_{ikj} = \Zi_{jki} = \Zi_{jik} = \Zk_{kij} = \Zj_{kji},\\
&\Zk_{ijk} = \Zj_{ikj} = \Zj_{jki} = \Zk_{jik} = \Zi_{kij} = \Zi_{kji}.
\end{align}
\end{subequations}

Like the second order tensors,  the third order tensors needed to build the velocity term $\Y$ are not independent from those to build $\X$.   More specifically, the expression for $\ng_{ijk}$, $\mg_{ijk}$, and all the $\ng$ and $\mg$ obtained by permutations of the indexes $ijk$, can be expressed as a linear combination of the four linearly independent vectors $\rg_{ijk}$, $\ug_{ijk}$, $\ug_{jki}$, $\ug_{kij}$. In turn, these vectors can be unequivocally expressed as a linear combination of the four linearly independent vectors $\Za_{ijk}$, $\Zi_{ijk}$, $\Zj_{ijk}$, $\Zk_{ijk}$. This particularly means that only four independent vectors  $\Za_{ijk}$, $\Zi_{ijk}$, $\Zj_{ijk}$, $\Zk_{ijk}$ are needed to compute 8 different vectors: $\rg_{ijk}$, $\ug_{ijk}$, $\ug_{jki}$, $\ug_{kij}$ (equation on $\X$) and  $\mg_{ijk}$, $\ng_{ijk}$, $\ng_{jki}$, $\ng_{kij}$ (equation on $\Y$), leading to the conclusion that equation on $\Y$ is not independent from the one on $\X$.

\subsubsection{Third-order trivial resonances}\label{sec:TR}

Trivial resonances are the consequence of the purely imaginary eigenspectrum $\pm i \omega_p$ of any conservative vibratory system. Then for two different indices $(i,j)$ one is always able to create a third-order trivial internal resonance relationship since $i\omega_j = +i\omega_i-i\omega_i+i\omega_j$. The mathematical consequence is that the treatment of the third order terms is complicated by taking them into account. The physical consequence is that in vibration theory, backbone curves are bent to create either hardening or softening behaviour~\cite{touze03-NNM,TouzeCISM}.

To better present the workaround that is needed to compute the third order tensors in case of a trivial resonance in a direct way, it is far more understandable to show how to compute them in modal basis first. Then we will extend the treatment to the direct computation. Even though trivial resonance can appear only when two indices are present instead of four (for a general cubic term from a fourth-order tensor), the presentation below is given for four different $(s,i,j,k)$. This allows to maintain generality in the derivation, as well as more simple coding since these calculations are typically run in nested loops.

Let us start by writing Eqs.~\eqref{eq:ZaZiZjZk} in modal basis and by component for a non-resonant row $s$:
\begin{subequations}
\begin{align}
&\modal{Z0}^s_{ijk}=((+\omega_i+\omega_j+\omega_k)^2-\omega_s^2)^{-1}\modal{P0}^s_{ijk},\\
&\modal{Z1}^s_{ijk}=((-\omega_i+\omega_j+\omega_k)^2-\omega_s^2)^{-1}\modal{P1}^s_{ijk},\\
&\modal{Z2}^s_{ijk}=((+\omega_i-\omega_j+\omega_k)^2-\omega_s^2)^{-1}\modal{P2}^s_{ijk},\\
&\modal{Z3}^s_{ijk}=((+\omega_i+\omega_j-\omega_k)^2-\omega_s^2)^{-1}\modal{P3}^s_{ijk}.
\end{align}\label{eq:ZaZiZjZk_modal_nonres}
\end{subequations}
In modal basis, mass and stiffness matrices reduce to simple scalars thanks to their diagonalisation, and the bar over the variables are simply removed. Let us now suppose that the $s$-th row is resonant, meaning that $\omega_s$ is equal to one of the combinations of $\omega_i$, $\omega_j$, and $\omega_k$.
Then, one of the terms at the denominator in Eqs.\eqref{eq:ZaZiZjZk_modal_nonres} will be zero and the inversion will no longer be possible. In such case, the terms $\modal{Z}^s_{ijk}$ (where again the simplified notation $\modal{Z}^s_{ijk}$  refers to $\modal{Z0}^s_{ijk}$ to $\modal{Z3}^s_{ijk}$) that aim at cancelling $\modal{A}^s_{ijk} + h^s_{ijk}$ and $\modal{B}^s_{ijk}$ must be set equal to zero, thus avoiding the singularity (problem known as small denominator), and the above mentioned $\modal{A}^s_{ijk}+ h^s_{ijk}$ and $\modal{B}^s_{ijk}$, that are no longer cancelled, must remain in the reduced dynamics.

In order to write this set of operations in compact form, one has to first define the four denominator matrices:
\begin{equation}
\modal{\mathbf{D}}=(\pm\omega_i\pm\omega_j\pm\omega_k)^2\I-\Ot,
\end{equation}
one for each combination of $\pm$. We leave it generic because this operation will apply to each $\mathbf{D}$.
One of these will be singular because its $s$-th row will be zero. In order to be able to invert all other nonzero rows while enforcing the constraint that the $s$-th row of $\z$ must be zero, 
one has to fulfil the following system:
\begin{equation}
\begin{cases}
&\mathbf{D}\z=\modal{\mathbf{P}}-\e_s P^s,\\
&\e_s^\text{T}\z = 0.
\end{cases}
\end{equation}
In this way, the $s$-th row of the first system is the identity $0=0$, the others satisfy Eqs~\eqref{eq:ZaZiZjZk_modal_nonres}, and the last equality ensures $\z^s=0$.
To express this system in compact form, the matrices must be augmented by the $\e_s$ vector so that the system becomes:
\begin{equation}
\begin{bmatrix}
\z
\\
P^s
\end{bmatrix}
=
\begin{bmatrix}
\modal{\mathbf{D}}&\e_s\\
\e_s^\text{T}&0
\end{bmatrix}^{-1}
\begin{bmatrix}
\modal{\mathbf{P}}
\\
0
\end{bmatrix} 
\end{equation}
For a non-resonant row labelled $p$, the system is coherent with Eqs.\eqref{eq:ZaZiZjZk_modal_nonres}; and for the resonant row labelled  $s$, the system enforces $Z^s=0$. One must notice that not only the matrix $\mathbf{D}$ has been augmented by a row and a column, but also the RHS vector is made of the force vector $\mathbf{P}$ plus a zero on its last row, as well as the solution vector will be one row longer than the sought vector $\z$, with the last row  being  the reduced dynamics term that cannot be cancelled.
The last step is then to express the same operation in physical basis. The equivalent of $\mathbf{D}$ in physical basis is $\mathbf{\bar{D}}=(\pm\omega_i\pm\omega_j\pm\omega_k)^2\M-\K$ and the equivalent of the vector $\e_s$ in physical basis is the vector $\M\phit_s$. In fact, the constraint on $\Z$ is expressed in physical basis by the mass orthogonality between $\Z$ and the $s$-th mode $\phi_s$. The expression for the generic $\Z$ in direct form finally reads:
\begin{equation}
\begin{bmatrix}
\Z
\\
P^s
\end{bmatrix}
=
\begin{bmatrix}
\mathbf{\bar{D}}  &  \M\phit_s\\
(\M\phit_s)^\text{T}&0
\end{bmatrix}^{-1}
\begin{bmatrix}
\modal{\mathbf{\bar{P}}}
\\
0.
\end{bmatrix}
\end{equation}

This augmentation procedure thus allows one to have a direct computation of all the needed trivially resonant third-order terms. Computationally speaking, since the evaluation of the $\z$ tensors is performed inside nested loops spanning over three indexes $i,j,k$, the augmentation procedure has to be actuated every time two of the three indexes are equal, say $i=j$, with the augmenting vector $\phit_s$ being the non equal remaining index $s=k$.

\subsection{Internal resonance}\label{sec:IR}

When deriving the theory of normal form, as already stated in the previous sections and fully detailed in~\cite{touze03-NNM,TOUZE:JSV:2006,TouzeCISM}, one has to take care of the occurrence of internal resonance. The general relationship for nonlinear resonance is found from Poincar{\'e} and Poincar{\'e}-Dulac's theorems. In the context of nonlinear vibratory systems composed of a purely imaginary eigenspectrum, it is customary to distinguish trivial internal resonance from non-trivial one. Third-order trivial resonances always exist. Consequently the normal form contains cubic-order terms that cannot be cancelled, see Eq.~\eqref{eq:ROM}. Each monomial term staying in the normal form is linked to a specific trivial internal resonance, and the previous section details how to handle these terms in the processing of the cubic terms.

Non-trivial internal resonances have not been taken into account yet since the main assumption holding from the beginning is that of a system free of these relationships. In case a third-order non-trivial internal resonance exist ({\em e.g.} a relationship of the form $\omega_p = \pm \omega_k \pm \omega_i \pm \omega_j$), then exactly the same procedure as the one explained in the previous section for trivial resonance has to be applied. This will let the corresponding cubic term of the nonlinear mapping to zero and make appear the corresponding monomial in the normal form. Consequently adapting the general reduced dynamics given by  Eq.~\eqref{eq:ROM} to a case with {\em e.g.} 1:1, 1:3 or a cubic combination resonance, is not difficult. Examples with 1:1 resonance can be found for example in~\cite{TOUZE:JSV:2006,TOUZE:JFS:2007}. Also the case of a three master modes ROM with 1:3 and 1:1 internal resonances will be highlighted in section~\ref{sec:beam}.

On the other hand, if second order internal resonance exists ({\em e.g.} 1:2 relationship $\omega_p = 2\omega_j$ or a combination resonance  $\omega_p = \omega_j \pm \omega_i$) then a special care has to be taken. Indeed Eq.~\eqref{eq:ROM} cannot be applied directly since the change in the second-order tensors will create new and important changes in the cubic terms. At present the method needs further refinement in order to compute fully the third-order tensor, so in this case it is advised to use only the  second-order DNF to create a ROM.

\subsection{Damping and forcing}\label{sec:dampforc}

In most of the cases one is interested in computing frequency-response curves of nonlinear structures using a dedicated ROM, and not only to predict the backbone curve of the unforced and undamped system. This section is devoted to explain how external forcing and viscous damping can be taken into account in the proposed reduction strategy using direct normal form, based on previous results already shown in~\cite{touze03-NNM,TOUZE:JSV:2006,TouzeCISM}.

For the external forcing, the strategy had already been proposed in~\cite{TOUZE:JSV:2006,TouzeCISM}. Since the reduced-order model used normal coordinates linked to the invariant manifolds that are tangent to their linear counterpart at origin, one can simply add the modal force at the right-hand side. This approximation does not lead to an exact solution since the correct formalism should take into account time-dependent manifolds. However, numerical results reported in \cite{Pesheck00,PesheckJSV,Jiang04,JiangForcing} clearly shows that the time-dependent variation of the manifold are negligible so that the results from unforced problem is meaningful. More recently, a mathematical proof that at the leading-order, this assumption of using a phenomenological forcing aligned with a curvilinear coordinates, is correct, has been given in~\cite{VERASZTO}, hence justifying again that this method can be used safely for including the forcing.

For the damping, general formulas for computing the normal form with modal damping have already been derived in~\cite{TOUZE:JSV:2006}. However, the coefficients are more complex to handle and a number of new terms appear also in the equations, see~\cite{TOUZE:JSV:2006} for a more detailed comparison between conservative and damped normal form. Hence deriving the general expressions for a damped system overshoot the mark of the present study and is postponed to further developments. However, in order to take into account in the ROM a meaningful damping term that aggregates the losses of all the slave modes, a simplified formula is given here and will be used in the simulation results presented in section \ref{sec:FRF}. The general idea of the present development is to restrict to the case of lightly damped systems, which is meaningful since geometric nonlinearities develop more easily in this context, whereas increasing damping generally favours linear behaviours.  Also, since the goal of the present work is to give a ROM strategy applicable for FE structures, a special emphasis is put on the case of Rayleigh damping which is currently used in FE codes, and general formula for a direct computation of new coefficients that take into account Rayleigh damping is derived.


The proposed strategy fully relies on the general results given in~\cite{TOUZE:JSV:2006} that are simplified and modified in order to tackle the problem at hand. More specifically, the proposed formula are given here only for the case of the second-order normal form. Indeed, computing the third-order terms in the nonlinear mapping with damping included needs a further development that needs to be accomplished in a further work. The main assumption is that of light damping. This means that,  in an asymptotic development relative to damping factors, only the leading order term is considered, thus simplifying the general formulas given in~\cite{TOUZE:JSV:2006}. 

The dynamics of the structure in FE nodes now reads:
\begin{equation}
\M\ddot{\X}+\C \dot{\X} +\K \X+\G(\X,\X)+\H(\X,\X,\X) = \zervec,
\label{eq:eom_phys_damped}
\end{equation}
where the only additional term as compared to Eq.~\eqref{eq:eom_phys} is $\C \dot{\X}$, and where the damping matrix $\C$ reads:
\begin{equation}
\C=\zeta_M \M + \zeta_K \K,
\end{equation}
following the definition of Rayleigh damping, with $\zeta_M$ and $\zeta_K$ two free parameters. If expressed in the modal basis, the equations of motion for a  generic mode $s$ reads:
\begin{equation}
\ddot{x}_s+
\zeta_s \dot{x_s} + 
\omega_s^2 x_s + \sum_{i=1}^N \sum_{j=1}^N g^s_{ij}x_i x_j +
 \sum_{i=1}^N \sum_{j=1}^N \sum_{k=1}^N h^s_{ijk}x_i x_j xk =0. 
\label{eq:eom_phys_damped}
\end{equation}
The parameter $\zeta_s$ can also be rewritten as  $\zeta_s=2 \xi_s \omega_s$ in order to make appear the  $\xi_s$ modal damping ratio. If $\C=\zeta_M \M + \zeta_K \K$, then $\zeta_s=\zeta_M + \zeta_K \omega_s^2$ and $\xi_s = (\zeta_M/\omega_s + \zeta_K \omega_s)/2$. Note that in~\cite{TOUZE:JSV:2006}, analytical asymptotic expansions for small damping are conducted by assuming the modal damping ratio to be small, $\xi_s \ll 1$.

In order to take into account the damping, three new tensors $\agd,\bgd,\cgd$ have to be added in the nonlinear mapping up to the second order, which now reads:
\begin{subequations}\begin{align}
&\X=
\sum^n_{i=1} \phit_i \R_i + 
\sum^n_{i=1}\sum^n_{j=1}
\ag_{ij}
\R_i\R_j+
\sum^n_{i=1}\sum^n_{j=1}
\bg_{ij}
\S_i\S_j+
\sum^n_{i=1}\sum^n_{j=1}
\cgd_{ij}
\R_i\S_j
,
\label{eq:nonlinear_change_x_phys_dam}
\\
&\Y=
\sum^n_i \phit_i \S_i + 
\sum^n_{i=1}\sum^n_{j=1}
\agd_{ij}
\R_i\R_j+
\sum^n_{i=1}\sum^n_{j=1}
\bgd_{ij}
\S_i\S_j+
\sum^n_{i=1}\sum^n_{j=1}
\cg_{ij}
\R_i\S_j
.
\label{eq:nonlinear_change_y_phys_dam}
\end{align}
\label{eq:nonlinear_change_phys_dam}
\end{subequations}


In the general case of arbitrary  damping values, all the tensors are modified, meaning that $\ag,\bg,\cg$ needs also to be recomputed in order to take properly into account the damping. However, as shown in~\cite{TOUZE:JSV:2006}, if light damping is considered, $\xi_s \ll 1$, then all the expressions can be simplified, and one finds that the leading-order term for the $\ag,\bg,\cg$ is the same as the one in the conservative case (the first adjustment due to damping being at order $\xi_s^2$). On the other hand,  $\agd,\bgd,\cgd$ involve only even powers of the damping in the asymptotics, so that the leading order is at order $\xi_s$ only: this term is thus now taken into account.



The expressions up to first order for the tensor $\cgd$ with the assumption of Rayleigh damping reads:
\begin{equation}\label{eq:cijdamping}
\cgd_{ij}=
(\zeta_M + 3\omega_i^2\zeta_K)\bg_{ij}
-(2\zeta_K)\ag_{ij}
+(-\zeta_M + 2\omega_i^2\zeta_K)(\Zss+\Zdd)
+(-\zeta_M + 2\omega_j^2\zeta_K)(\omega_i/\omega_j)(\Zss-\Zdd)
\end{equation}
with $\ag_{ij}$ and $\bg_{ij}$ given by Eq.~\eqref{eq:abcphys} (conservative values), and the new terms as:
\begin{subequations}\label{eq:ZssZdd}
\begin{align}
\Zss_{ij}=((+\omega_i+\omega_j)^2\M-\K)^{-1}\M\, \Zs_{ij},\\
\Zdd_{ij}=((-\omega_i+\omega_j)^2\M-\K)^{-1}\M\, \Zd_{ij}.
\end{align}
\end{subequations}
As one can observe, the structure of the equations are close to the undamped case, the difference being that $\Zss_{ij}$ and $\Zdd_{ij}$ are more involved, but implying the same kind of operations as $\Zs_{ij}$ and $\Zd_{ij}$. The full expression for $\cgd_{ij}$ clearly underlines that it is a first-order term on the  damping coefficients. Interestingly, only the damping coefficients of the master modes $i,j$ appear in Eq.~\eqref{eq:cijdamping}, meaning in fact that the assumption of small damping needs only to be retained for the master modes. This is  particularly significant since ROMs are generally built for low-frequency modes that have the smallest damping ratios. It underlines that the assumption of light damping should not be too restrictive since slave modes can have larger damping ratios. 

From Eq.~\eqref{eq:cijdamping} defining $\cgd_{ij}$, the computation of the two other second-order tensors simply reads:
\begin{subequations}\label{eq:albetdamp}
\begin{align}
&\agd_{ij}=-\omega_i^2 \cgd_{ij},\\
&\bgd_{ij}=\cgd_{ij}-(\zeta_i+\zeta_j)\bg_{ij}.
\end{align}
\end{subequations}
To conclude on the nonlinear mapping, in case of damping, one simply has to keep the $\ag,\bg,\cg$ as in the conservative case, and compute the three new tensors $\agd,\bgd,\cgd$ from \eqref{eq:cijdamping}-\eqref{eq:albetdamp}.

The reduced-order dynamics is also modified by the damping terms. Following \cite{TOUZE:JSV:2006}, and restricting to the case of the second-order  normal form, the reduced dynamics now reads:
\begin{equation}
\ddot{\R}_r+ \left( \zeta_M +  \zeta_K \omega_r^2 \right) \dot{\R}_r    + \omega^2_r\R_r+
\sum^n_{i=1}\sum^n_{j=1}\sum^n_{k=1}
\left[ (A^r_{ijk} +   h^r_{ijk})  \R_i\R_j\R_k
+
B^r_{ijk}\R_i\dot{\R}_j\dot{\R}_k
+
C^r_{ijk}\R_i\R_j\dot{\R}_k \right]
=0.
\label{eq:ROM_SO_damped}
\end{equation}
Apart from the linear damping term, the proposed strategy to take damping into accounts leads to the appearance of the term $C^r_{ijk}\R_i\R_j\dot{\R}_k$. The coefficient $C^r_{ijk}$ is a summation on all the modes including the slave ones. This added term can thus be understood as an aggregate nonlinear damping that takes into account the damping of all the slave modes in order to better approximate the losses on the invariant manifold. From~\cite{TOUZE:JSV:2006}, the new coefficient reads:
\begin{equation} \label{eq:C}
C_{ijk}^r = \sum_{s=1}^{N} 2\,g_{is}^r c_{jk}^s ,
\end{equation}
The last needed formula is the equivalent of Eq.~\eqref{eq:ABphys} within the direct formulation in order to compute $C_{ijk}^r$, which is analogous to  the fourth-order tensors  $\At$ and $\Bt$, in order to avoid the full computation of all the $g^r_{rs}$ appearing in \eqref{eq:C}.  After the same kind of  manipulations as those reported in section~\ref{sec:NFo2}, explicit expression for $C^r_{ijk}$ finally reads:
\begin{equation}
C^r_{ijk} = 2\, \phit_r^\text{T}\, \G(\phit_i,\cgd_{jk}),
\end{equation}
hence allowing to express all the needed quantities so that they can be computed in a direct way from the FE discretisation. 

Note that all other terms appearing in the reduced dynamics are left unchanged as compared to \eqref{eq:ROM_SO}. In the case of a single master coordinates labelled $p$, then only one extra term has to be taken into account in the reduced dynamics, $C^p_{ppp}\R_p^2\dot{\R}_p$. In case of more than one master coordinate, a simplified approach could have been to consider only the $C^p_{ppp}$ terms in the reduced dynamics (\emph{e.g.} in case of two master modes $1$ and $2$, only $C^1_{111},C^2_{222}$ could have been added). However the cross couplings terms have been found to be important in the reduction process, this will be illustrated in Section~\ref{sec:FRF}.

\subsection{Practical implementation and algorithm}\label{sec:algo}

In this section, a simplified overview of the computational algorithm is given, together with further considerations on the implementation and the non-intrusiveness of the method. In order to simplify the presentation, only the treatment of the conservative terms is made explicit. Adding the damping terms can be done easily following the guidelines given in the previous section.

In the present framework of distributed geometric nonlinearity where all the oscillator equations are fully coupled with quadratic and cubic terms, direct extractions from the finite element model of the full tensors $\G$ and $\H$  is not feasible. The proposed method is direct in the sense that 
 only the nonlinear force tensors $\G(\bm{\phi}_i,\bm{\phi}_j)$ and $\H(\bm{\phi}_i,\bm{\phi}_j,\bm{\phi}_k)$ are required with $i,j,k\in[1,n]$ only spanning the master modes.

These can be easily obtained thanks to the known STEP method, that computes the quadratic and cubic forces generated in the structure when a displacement along a combination of eigenvectors is imposed, see {\em e.g.}~\cite{muravyov,givois2019}.  
The number of STEP method calculations is then equal to $n(n+1)(n+2)/6$ (scales with $n^3$) for the $\H$ tensor and $n(n+1)/2$ (scales with $n^2$) for the $\G$ tensor if the symmetry property of the tensors are exploited.
 
Also one of the main advantage of the proposed method is to rely on a direct computation solely involving  a selection of a priori master coordinates from usual physical considerations (on the nature of the forcing or the dynamics at hand). Since the ROM is directly expressed in the invariant-based span of the phase space, the curvatures induced by the nonlinear mapping allows to directly compute the important, non-resonant couplings. There is no extra need to seek the slave modes coupled to the master ones by a convergence study: they are directly embedded in the nonlinear mapping and thus taken into account. The consequence is that there is no need to compute the full eigenvector matrix $\V$, only  the master modes are needed and stored in the matrix $\EigVec$.

A simplified overview of the algorithm is shown in Alg.~\ref{fig:alg}. It relies on four main separate functions: \texttt{Eig}, \texttt{StEP}, \texttt{Calc$\_$O2} and \texttt{Calc$\_$O3}. \texttt{Eig} is the usual eigensolver and is not explicated further. \texttt{StEP} implements the StEP method following the general and known guidelines, see for example~\cite{muravyov,Perez2014,givois2019}. The starting point of the algorithm is given by the user, with the selection of the master modes of interest, from which the matrix $\EigVec$, together with their companion eigenfrequencies, are formed. The STEP method is then applied in order to derive the quadratic and cubic terms of the master modes of interest.

\begin{algorithm}[h!]
\caption{Direct Normal Form}\label{fig:alg}
\SetAlgoLined
\DontPrintSemicolon
\SetKwFunction{Eig}{Eig}
\SetKwFunction{StEP}{StEP}
\SetKwFunction{SOT}{Calc$\_$O2}
\SetKwFunction{TOT}{Calc$\_$O3}
\SetKwInOut{Input}{Input}
\SetKwInOut{Output}{Output}
\SetKwInOut{OptOutput}{OptionalOutput}
\SetKwInOut{Functions}{Functions}
\Input{$\K,\M$}
\Functions{\Eig, \StEP, \SOT,\TOT}
\Output{$\EigVec, \EigVal,\ag,\bg,\cg, \At, \Bt,\h$}
\OptOutput{$\rg,\ug,\mg,\ng$}
\BlankLine\BlankLine
\emph{Compute eigenproblem only for the master modes}\;
$\phit_r, \omega_r\leftarrow$ \Eig($\K,\M$)\;
\BlankLine\BlankLine
\emph{Compute StEP method on the master modes $\EigVec$}\;
$\G(\phit_i,\phit_j),\H(\phit_i,\phit_j,\phit_k)\leftarrow$ \StEP($\phit_i,\phit_j,\phit_k$)\;
\BlankLine\BlankLine
\emph{Compute second order mapping tensors $\ag,\bg,\cg$}\;
$\ag_{ij},\bg_{ij},\cg_{ij}\leftarrow$\SOT($\K,\M,\omega_i,\omega_j,\G(\phit_i,\phit_j)$)\;
\BlankLine\BlankLine
\emph{Compute StEP method on $\EigVec,\ag$ and $,\EigVec,\bg$}\;
$\G(\phit_i,\ag_{jk})\leftarrow$ \StEP($\phit_i,\ag_{jk}$)\;
$\G(\phit_i,\ag_{jk})\leftarrow$ \StEP($\phit_{i},\bg_{jk}$)\;
\BlankLine\BlankLine
\emph{Obtain third order forces tensors $\Ag$ and $\Bg$}\;
$\Ag_{ijk}\leftarrow 2\, \G(\phit_i,\ag_{jk}) $\;
$\Bg_{ijk}\leftarrow 2\, \G(\phit_i,\bg_{jk})$\;
\BlankLine\BlankLine
\eIf{DNF up to Second Order}{
\emph{Compute full third order reduced dynamics tensors}\;
$h^r_{ijk}\leftarrow \phit_r^\text{T}\H(\phit_i,\phit_j,\phit_k)$\;
$\modal{A}^r_{ijk}\leftarrow \phit_r^\text{T}\Ag_{ijk}$\;
$\modal{B}^r_{ijk}\leftarrow \phit_r^\text{T}\Bg_{ijk}$\;
\BlankLine\BlankLine
}{
\emph{Compute third order mapping tensors}\;
${\rg}_{ijk},{\ug}_{ijk},{\mg}_{ijk},{\ng}_{ijk}\leftarrow$
\TOT(${\K},{\M},\omega_i,\omega_j,\omega_k,{\Ag}_{ijk},{\Bg}_{ijk},\H(\phit_i,\phit_j,\phit_k)$)\;
\BlankLine\BlankLine
\emph{Fill third order reduced dynamics tensors solely with trivially resonant terms}\;
$h^r_{ijk}\leftarrow \phit_r^\text{T}\H(\phit_i,\phit_j,\phit_k)$\;
$\modal{A}^r_{ijk}\leftarrow \phit_r^\text{T}\Ag_{ijk}$\;
$\modal{B}^r_{ijk}\leftarrow \phit_r^\text{T}\Bg_{ijk}$\;
}
\end{algorithm}

The operations contained in the functions \texttt{Calc$\_$O2} and \texttt{Calc$\_$O3} have been already detailed in Sec.~\ref{sec:NFo2} and Sec.~\ref{sec:NFo3}, but their practical implementation is yet to be discussed.  For the sake of simplicity, the discussion is here restricted to the case of a single master mode, but of course the same considerations also apply to the more general case of multiple master modes.

The outputs of \texttt{Calc$\_$O2} are the $\ag$, $\bg$, $\cg$ second-order tensors. As shown in Section~\ref{sec:NFo2}, they are derived from the computation of internal variables $\Zs$ and $\Zd$. Let us detail this computation and comment its effectiveness with regard to a non-intrusive method, in the case of a single master mode $i$.  Eqs.~\eqref{eq:ZsZd_single} gives the expressions of $\Zs$ and $\Zd$, which are written in their explicit form for a direct solution, i.e. with the unknown vectors alone on the left hand side. In the actual implementation, there is however no need to perform a matrix inversion because, in the FE solver, a linear algebraic system is solved instead:
\begin{subequations}
\begin{align}
&((2\omega_i)^2\M-\K)\Zs=\G(\phit_i,\phit_i),
\label{eq:Zs_sys}\\
&\K\,\Zd=-\G(\phit_i,\phit_i).
\label{eq:Zd_sys}
\end{align}
\end{subequations}
In  Eq.~\eqref{eq:Zd_sys}, the linear system can be solved directly by  performing a linear static analysis that computes the unknown displacement vector $\Zd$ when the structure is subjected to the force vector $-\G(\phit_i,\phit_i)$. This type of analysis is a standard operation in every FE software and its computational cost is quite small even for very large models. The same kind of operation has already been used in the context of SMD and is considered as a non-intrusive calculation. The linear system in Eq.~\eqref{eq:Zs_sys} is a bit different and involves a linear combination of $\K$ and $\M$, but without an increase in the size of the system to solve. Despite this operation is not a standard one in a FE software, its computational cost is again quite low. To be able to perform this operation, the FE software must allow the user to script and to do matrix operations online. In this way, the $\K$ and $\M$ matrices are never exported from the software but only the resulting vector is.

The next steps of the algorithm is then to compute the $\Ag$ and $\Bg$ tensors following Eqs.~\eqref{eq:ABphys}. This is performed in lines 8 and 9 of the algorithm, by using the STEP function. Indeed, one can note that STEP is a calculation method, that can be applied with any input vectors needed. Even though, in its first derivation given in~\cite{muravyov}, it was only thought for a non-intrusive computation of the modal nonlinear coupling coefficients, so that the entries of the procedure were prescribed displacements along given eigenmodes, the procedure can also be used more generally, as done here with entries composed of one eigenvector and a vector $\ag_{ij}$ or $\bg_{ij}$. Once $\Ag$ and $\Bg$ computed, their counterpart $\At$ and $\Bt$ are found by using a projection involving $\EigVec$, which is needed for the reduced-order dynamics.

The last part of the algorithm distinguishes if the user only needs the second-order normal form or the third-order. In case the third-order is needed, then the function \texttt{Calc$\_$O3} is called, producing the computations explained in Section~\ref{sec:NFo3}.

For all our computations, the open software Code\_Aster \cite{ASTER} has been used. All the calculations have been simply implemented using external scripts driving the master code, without the need of entering intrusively in the code.

\section{Numerical results}\label{sec:results}

\subsection{A clamped-clamped beam with internal resonance}\label{sec:beam}

The first  example is a clamped-clamped straight beam, for which the amplitude-frequency backbone curves will be computed with the direct normal form approach. This example has interesting features since it presents, at high levels of vibration amplitudes, internal resonances between the nonlinear frequencies of the system. This creates resonance loops in the backbones where strong interactions between the modes exist, see {\em e.g.}~\cite{Lewandowki97a,KerschenNNM09,PeetersNNM09}. This salient feature shares common points with internal resonance between the eigenfrequencies of the system, already discussed in section~\ref{sec:IR}, for which special care needs to be taken since corresponding to a resonant monomial term in the normal form. In the present case, the resonance occurs between the nonlinear frequencies, at large amplitudes.  Such a beam example has also been studied recently in~\cite{SOMBROEK2018}, where these resonance loops have been reported in Frequency-Energy Plots (FEP). Consequently the results shown in~\cite{SOMBROEK2018} will serve here as a guideline in order to detect if the same behaviour can be retrieved with the DNF.

\begin{figure}[h!]
\centering
\includegraphics[height=4cm]{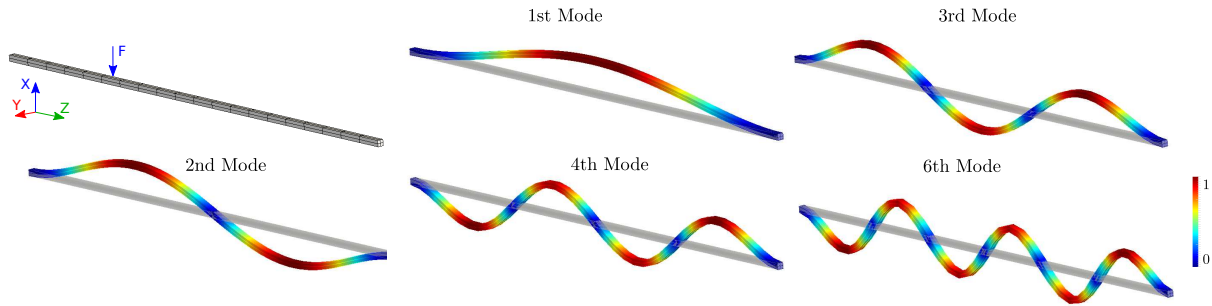}
\caption{Mesh selection and geometry of the straight clamped-clamped beam, as well as eigenmode shapes corresponding to bending modes number 1 to 4 and mode 6. The bending vibration direction is along $x$ while the in-plane direction is $z$.}
\label{fig:beam_mesh}
\end{figure}

The model in~\cite{SOMBROEK2018} used simple 1D beam elements thus resorting to a simplified kinematics, and 30 elements gave the space discretisation. In order to cope with a more realistic mesh, closer to general assumptions of elasticity, 3D block elements with three displacements per nodes, have been used for meshing the clamped-clamped beam. The FE model   consists of a mesh of 80 hexahedrons (20 along the axis, 2x2 in the section) with 20 nodes each, for a total number of 621 nodes, and 1863 degrees-of-freedom (dofs). The dimensions of the beam are $L=1$ m along the $z$ axis, $b=h=0.01$ m. The material properties are the following: $E=210$ GPa, $\rho=8750$ kg/m$^3$, $\nu=0.3$.  The mesh is shown in Fig.~\ref{fig:beam_mesh} together with the eigenmodeshapes of modes 1 to 4 and 6. In the rest of the paper, the displacement along $x$ will be denoted as $u$, while the displacement along the in-plane direction $z$, is denoted as $w$. No motion is allowed along $y$.

The six lowest eigenfrequencies of the beam, corresponding to bending modes, are  reported in the first row of Table~\ref{tab:freq_beam}. In the second row, the frequency ratio with respect to mode 1 is given, showing that the eigenfrequency of mode 3 is a bit larger than 5 times the first. At the linear level, a 5:1 internal resonance does not exist. However, since the beam has a hardening behaviour, increasing the amplitude will make the fulfilment of 5:1 ratio possible, so one could expect for the first NNM, a strong interaction in 5:1 ratio with mode 3, a result that has  indeed already been reported in~\cite{SOMBROEK2018}. The last row of Table~\ref{tab:freq_beam} shows the frequency ratios with respect to mode 2. One can see that a 3:1 internal resonance may be excited once the hardening behaviour has sufficiently increased the nonlinear frequencies. So a 3:1 internal resonance loop with an interaction with mode 4 can be expected for NNM 2. As reported in~\cite{SOMBROEK2018}, this will be indeed the case, as well as, at larger amplitudes, a 5:1 resonance with mode 6.

\begin{table}[h!]\centering\small
\begin{tabular}{l| r r r r r r}
\rule{0pt}{13pt}Mode Number  &
1&   2&   3&   4&   5& 	6\\
\hline
\rule{0pt}{13pt}Frequency (Hz)& 
50.900&		140.74&		277.09&		460.64&		692.93&		975.85\\
Frequency Ratio with mode 1& 
1&   &   5.44&    &   &\\
Frequency Ratio with mode 2& 
 &   1&      &   3.27 &   &6.93\\
\end{tabular}
\caption{Linear modes frequencies and frequency ratios  with mode 1 and 2.}
\label{tab:freq_beam}
\end{table}
%
%

\begin{figure}[h!]
\centering
\includegraphics[height=9cm]{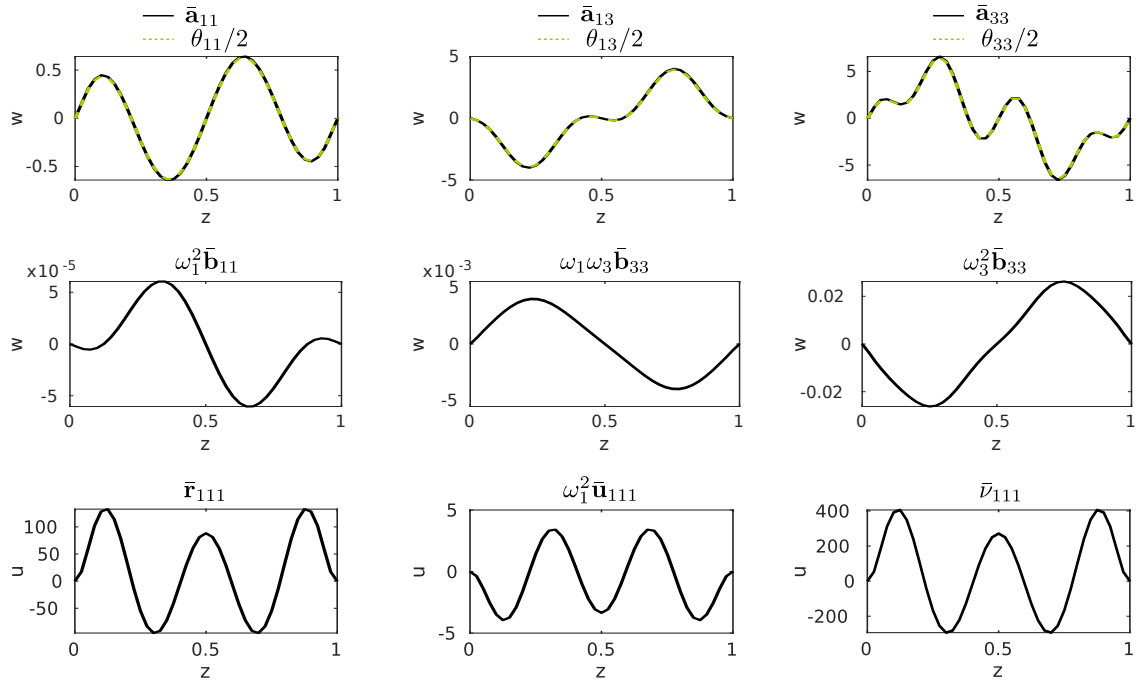}
\caption{Illustration of the physical content of some of the quadratic and cubic tensors used in DNF approach. First line: $\ag_{11}$, $\ag_{13}$ (black lines) compared to static modal derivatives $\theta_{11}$, $\theta_{13}$ and $\theta_{33}$ (dashed green line). Second line: quadratic vectors $\bg_{11}$, $\bg_{13}$ and $\bg_{33}$. Third line: cubic vectors:  $\rg_{111}$, $\mg_{111}$ and $\ng_{111}$.}\label{fig:beam_tensors}
\end{figure}

Fig.~\ref{fig:beam_tensors} shows some of the components of the quadratic and cubic tensors used in the nonlinear mapping, Eq.~\eqref{eq:nonlinear_change_phys03}. The first line shows quadratic $\ag_{ij}$ vectors for some specific combinations: $\ag_{11}$, $\ag_{13}$, and $\ag_{33}$, and give also a direct comparison with the corresponding static modal derivative (SMD) $\theta_{11}$, $\theta_{13}$ and $\theta_{33}$.  As explained in section~\ref{sec:NFo2}, in a simplified case where a slow/fast assumption can be assumed, then $\ag_{ij}$ vectors will point to the same direction as static modal derivatives, and $\bg_{ij}$ vectors should be negligible. This property is indeed verified in this example: all three selected $\ag_{ij}$ examples are exactly the same as their SMD counterparts. Interestingly, $\ag_{ij}$ vectors are able to recover the most important quadratic couplings between eigenmodes. As analysed {\em e.g.} in~\cite{Vizza3d}, for flat structures such as the straight beam considered, quadratic couplings arise only in the oscillator equations governing the in-plane modes dynamics, and involve two bending modes. With this respect, one understands why the  $\ag_{ij}$ vectors only involve in-plane motions, such that their main component is along $w$ ($z$ direction) in Fig.~\ref{fig:beam_tensors}. Also,   slow/fast assumption in order to use safely SMD has been estimated in~\cite{Vizzaccaro:NNMvsMD} as soon as the ratio between slave and master mode is larger than 4. Here the ratio between the first bending mode and the fourth axial mode is approximately $192.4$. Second, $\ag_{11}$ allows to directly recover the fact that for the first bending mode of a beam, the most important coupling is with the 4th axial mode, see {\em e.g.}~\cite{givois2019,Vizza3d,YichangICE} for numerical examples highlighting this result. It is thus fully logical to observe that $\ag_{11}$ have the shape of this fourth axial. The direct consequence is that there is no need to analyse beforehand the mode couplings, since they are automatically embedded in the added quadratic tensors.  Finally, as shown in~\cite{givois2019}, bending mode 3 has a strong quadratic coupling with in-plane modes number 2, 6 and 8. Consequently,  $\ag_{33}$  shows an axial deformation being a combination of these three mode shapes.

The second line of Fig.~\ref{fig:beam_tensors} shows the same plot as the first line but now for the  $\bg_{ij}$ vectors: $\bg_{11}$, $\bg_{13}$ and $\bg_{33}$.
These additional vectors are not taken into account if one uses the quadratic manifold from SMD. In this specific case of a straight beam where the slow/fast assumption is very well fulfilled, one can observe that the $\bg_{ij}$ vectors have, as anticipated, very small amplitudes, and can thus be safely neglected. These first results, highlighting how the nonlinear couplings are embedded into the additional vectors of the second-order term in the mapping, explains why in this simple case of a straight beam, the results provided by SMD are very good and allows to retrieve correct predictions, as reported in~\cite{SOMBROEK2018}. However, adding curvature to the structures and thus more complex couplings between modes and disappearance of the slow/fast assumption, it may be anticipated that SMD will not produce correct results anymore, as shown {\em e.g.} in~\cite{Vizzaccaro:NNMvsMD}. The reflection of that fact should then appear   in the behaviours of $\bg_{ij}$ vectors. This specific illustration is reported to a future work but preliminary comparisons are already reported in~\cite{Vizzaccaro:NNMvsMD}.

Finally, the third line of Fig.~\ref{fig:beam_tensors} shows three of the third-order vectors: $\rg_{111}$, $\mg_{111}$ and $\ng_{111}$. Again, for a flat structure, cubic couplings involve only bending modes. Consequently, these correction vectors are in the $x$ direction and involve mostly bending displacement $u$. Their shape also underlines the fact that, due to symmetry reasons, a clear separation exist between odd and even modes, the two families being coupled together with no cross-coupling between them. Consequently, $\rg_{111}$, $\mg_{111}$ and $\ng_{111}$ shows only combinations of odd modes and underlines why mode 1 couples only with modes 3, 5, and so on, and the same for mode 2 with mode 4, 6 and so on.

\begin{figure}[h!]
\centering
\includegraphics[height=6cm]{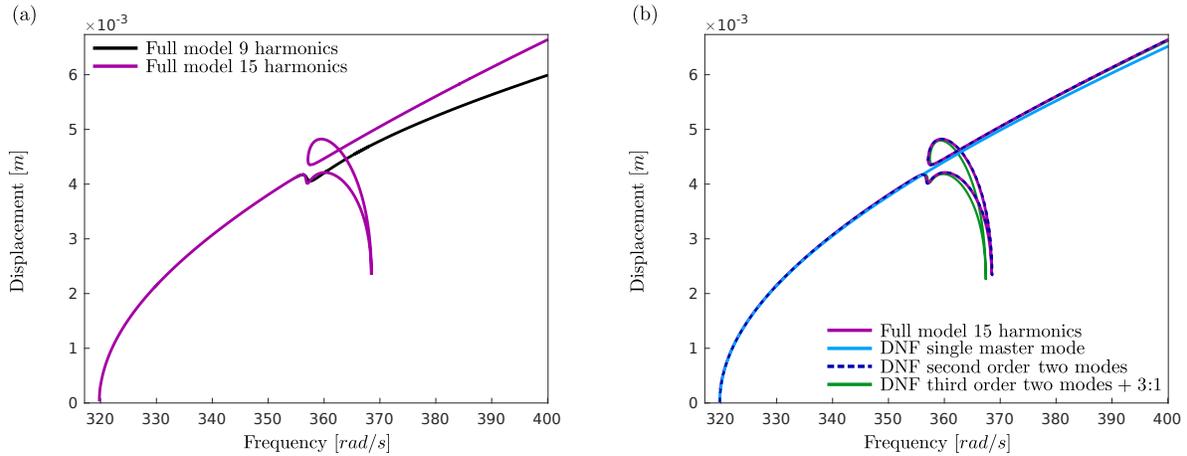}
\caption{Amplitude-frequency backbone curve for the first NNM of the clamped-clamped beam. (a) reference full-order solution obtained with 9 harmonics (black line) and with 15 harmonics (purple). Stability is not reported. (b) comparison of the reference solution to different ROMs computed with the direct normal form (DNF). Light blue curve: direct normal form with a single master coordinate corresponding to mode 1. Dashed blue line: second-order direct normal form with two master coordinates 1 and 3, green line: third-order direct normal form with two master modes (1 and 3) and inclusion of the resonant monomials corresponding to 3:1 internal resonance.}
\label{fig:beamNNM1}
\end{figure}

We now turn to the analysis of the backbone curves for the first and the second mode. In each case, a reference, full-order simulation, is performed thanks to numerical continuation on all the degrees of freedom of the structure. A parallel implementation using harmonic balance method and pseudo arc-length continuation, as reported in~\cite{Blahos2020}, is used. In this beam case, due to some limitations of this algorithm, the beginning of the backbone is numerically found by using a small amount of harmonic forcing. Then, as soon as the first points are found, forcing is removed and continuation restarted.

Fig.~\ref{fig:beamNNM1}(a) shows the reference result for the first mode obtained with the full-order model. As anticipated, a resonance loop appears along the backbone due to the excitation of the 5:1 internal resonance between mode 1 and mode 3. Interestingly, two results from the computation of the full-order model are shown in  Fig.~\ref{fig:beamNNM1}(a). The first one has been obtained by using 9 harmonics in the numerical solution for continuation. As observed, this solution is not converged since the complete resonance loop is partially missed, in contrast to the converged solution obtained with 15 harmonics. This result underlines that in the resonance loops, complex high-order interactions are activated between the harmonics of the solution and a too crude truncation oversimplify the analysis and misses important detail of the nonlinear dynamics.

In Fig.~\ref{fig:beamNNM1}(b), the reference solution is compared to  different ROMs computed with the direct normal form (DNF). The first ROM is obtained by retaining a single master coordinate in the reduction strategy, corresponding to mode 1 (DNF single master mode, light blue curve). As it could be expected, the single-mode solution offers a drastic reduction, is able to recover the correct hardening behaviour of the beam, but is not able to retrieve the internal resonance, since a strong interaction with mode 3 is activated. The minimal reduced model should consists of at least two master coordinates. Consequently different ROMs including two master coordinates corresponding to mode 1 and 3 are tested.

The second ROM is computed thanks to the second-order normal form presented in section~\ref{sec:NFo2}, and two master coordinates (modes 1 and 3). The main advantage of this strategy is that all possible combinations of higher-order resonance (from order three) have not been dealt with the nonlinear mapping. Consequently they are still present in the cubic terms and can express themselves if needed. This strategy is particularly meaningful in the case considered since a 5:1 internal resonance is at hand, which would formally need to push all the normal form development up to order five, which is beyond the scope of the present study. The ROM obtained with two master modes and second order normal form is able to properly retrieve the 5:1 internal resonance, without any extra effort from the analysis. This is explained by the fact that the second-order resonances have been correctly treated, but then stopping at this order let all other possibilities free. 

On the other hand, using the third-order normal form for getting back this 5:1 relationship opens the doors to new questions. In the best setting, one should go to order five and add only the resonant monomial terms corresponding to the 5:1 relationship. Since this calculation is far more difficult, different options are considered. The first, already commented option, is to stop at second order, leaving all higher-order terms there. This option gives very good result in this case. Its main drawbacks are twofold. First an order of accuracy in the nonlinear mapping has been lost as compared to what can be easily computed, meaning that the approximation of the underlying invariant manifold is less accurate. Second, the number of cubic terms in the ROM is more important, which is not a big issue since only two modes are considered. A second option would be to select two master coordinates and use the normal form up to the third order, using the solution obtained without considering internal resonance, see Eq.~\eqref{eq:ROM} for the reduced dynamics. In this case the same solution as with a single master coordinate will be retrieved, because no invariant-breaking terms are present in Eq.~\eqref{eq:ROM}. 


A third option consists in adding invariant-breaking terms in order to excite the coupling and retrieve the 5:1 internal resonance loop with only cubic terms. Since internal resonance appears due to nonlinear interactions between harmonics of the solution, it is meaningful to add invariant-breaking terms related to lower-order internal resonance, in this case 3:1 and 1:1. In the first case (3:1 resonance), two terms are added to the dynamics due to 3:1 resonance: a $R_1^2 R_3$ term for the first oscillator (and all their ancillary terms involving velocities), and a $R_1^3$ on the second master coordinate (corresponding to mode 3, again with all its companion terms with velocities). The dynamics of these ROMs are fully explicated in Appendix \ref{app:romresdet} so that the reader can get a better understanding of the equations used. 

The resulting backbone is shown in Fig.~\ref{fig:beamNNM1}(b), and is named DNF third order + 3:1. Interestingly, this ROM can recover the internal resonance loop of the 5:1, with only a slight departure from the reference solution. Finally if one also considers the terms due to 1:1, it means adding the $R_3^3$ term on $R_1$ equation, and $R_1 R_3^2$ monomial on $R_3$ equation (see Appendix \ref{app:romresdet} for details). In this case the model is simply the same as the one obtained with second-order normal form, since all the cubic terms are now present. So this case is equivalent to the result already obtained with the second-order normal form, and gives a perfect match.

\begin{figure}[h!]
\centering
\includegraphics[height=6cm]{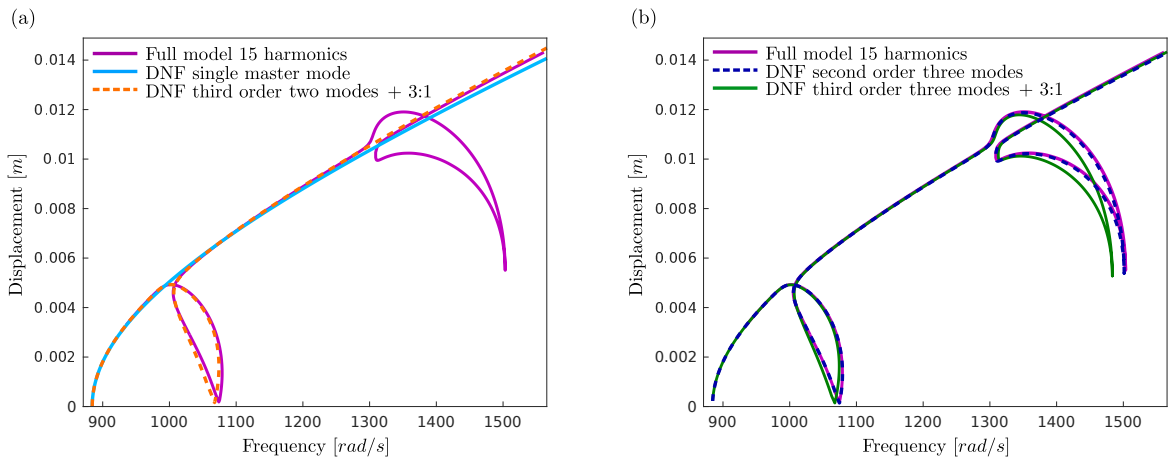}
\caption{Amplitude-frequency backbone curve for the second NNM of the clamped-clamped beam. (a) Reference solution obtained with 15 harmonics (purple) compared to ROM considering only master coordinate of mode 2 (DNF single master mode, light blue), and a ROM with the third-order normal form with modes 2 and 4 and the resonant monomials corresponding to 3:1 internal resonance (DNF third-order + 3:1, dashed red curve). (b) Reference solution is compared to a ROM obtained with the second-order normal form with 3 master coordinates corresponding to modes 2, 4 and 6 (dashed blue), and to a ROM obtained with the third-order normal form, 3 master coordinates, and resonant monomials corresponding to 3:1 between mode 2 and 4 and 2 and 6.}
\label{fig:beamNNM2}
\end{figure}

Fig.~\ref{fig:beamNNM2} shows the  results obtained for the second NNM. As expected from the linear  analysis and from the results already reported in~\cite{SOMBROEK2018}, the full-order solution shows two successive loops, corresponding to the fulfilment of 3:1 internal resonance between the nonlinear frequencies of mode 2 and mode 4, and the 5:1 resonance between mode 2 and mode 6. Again, different reduced-order models are compared in order to better understand their capabilities. A first ROM is built using a single master coordinate corresponding to mode 2, and compared to full-order solution in Fig.~\ref{fig:beamNNM2}(a).  This ROM allows to predict the correct hardening behaviour, but cannot catch the resonance loops since the resonant dynamics is intrinsically of higher dimension.  A second ROM with two master coordinates corresponding to mode 2 and 4 is then built, using the third-order normal form. Since a 3:1 internal resonance is expected, the resonant monomials corresponding to 3:1 resonance between modes 2 and 4 are taken into account in this reduced dynamics. The result is shown in Fig.~\ref{fig:beamNNM2}(a). It underlines that this ROM is capable of reproducing the loop of 3:1 internal resonance with a fair accuracy, but of course is not able to catch the second loop since a nonlinear interaction with mode 6 comes into play.

Fig.~\ref{fig:beamNNM2}(b) compares the results provided by two ROMs composed of three master coordinates corresponding to modes 2, 4 and 6. The first one is computed with the second-order normal form. As in the case of the first NNM, it gives a perfect match to the reference solution and is able to recover the two loops of internal resonances corresponding to 3:1 resonance with mode 4 and 5:1 resonance with mode 6. Note that in some sense this ROM is minimal since only three master coordinates have been used. To our point of view, it is not possible to reproduce such backbones with less than three master coordinates since nonlinear interactions are between mode 2 and 4 and then between mode 2 and 6. On the other hand, the same behaviour has been successfully reported in~\cite{SOMBROEK2018}, but their model were composed of 9 modes (3 linear modes plus 6 static modal derivatives).

The last ROM tested still contains three master coordinates and applies the third-order normal form. Resonant monomial corresponding to 3:1 resonance between mode 2 and 4 have been added to reproduce the first resonance loop. For the second loop implying a 5:1 resonance, the same problem appears as with the case of NNM 1: a perfect recovering of this resonance with the normal form theory would need to push the asymptotic development up to order five. Consequently, in the line of the previous study on NNM 1, only the resonant monomials corresponding to a 3:1 resonance between mode 2 and 6 have been added. This ROM contains far less nonlinear terms in the reduced dynamics than the second-order normal form. It is able to well reproduce the two resonance loops, however with a small difference as compared to the full-order solution. 

These examples clearly underline that selecting the second-order normal form is generally a very good option when odd resonances are present. Indeed, comparing in this case the reduced dynamics given by second and third-order DNF shows that the only difference is that the non-resonant terms have been removed in the dynamics for the third-order. On the other hand, the higher-order corrective terms that should come into play since the computation of normal form is intrinsically an asymptotics, have not been taken into account, based on the fact that the dynamics is truncated up to order three. This means that in the present version, the second-order normal form contains more information and is capable of reproducing higher-order resonance as shown here with the 5:1 case. The main drawback is that in the reconstruction process, the nonlinear mapping is less accurate since truncated to order two. But the gain in the reduced dynamics, which contains much more terms, seems to be more important.

\subsection{FE model of a fan blade}\label{sec:blade}

In this section, a FE model of a fan blade is selected in order to test the method with an application of high interest in engineering, with a more complex geometry and large-amplitude vibrations exciting geometric nonlinearities. The backbone curve of the first mode of the structure will be computed with the DNF approach and compared to a reference solution obtained from direct continuation of all the dofs of the structure.

\begin{figure}[h!]
\centering
\begin{subfigure}{.32\textwidth}\centering
\includegraphics[height=4.5cm]{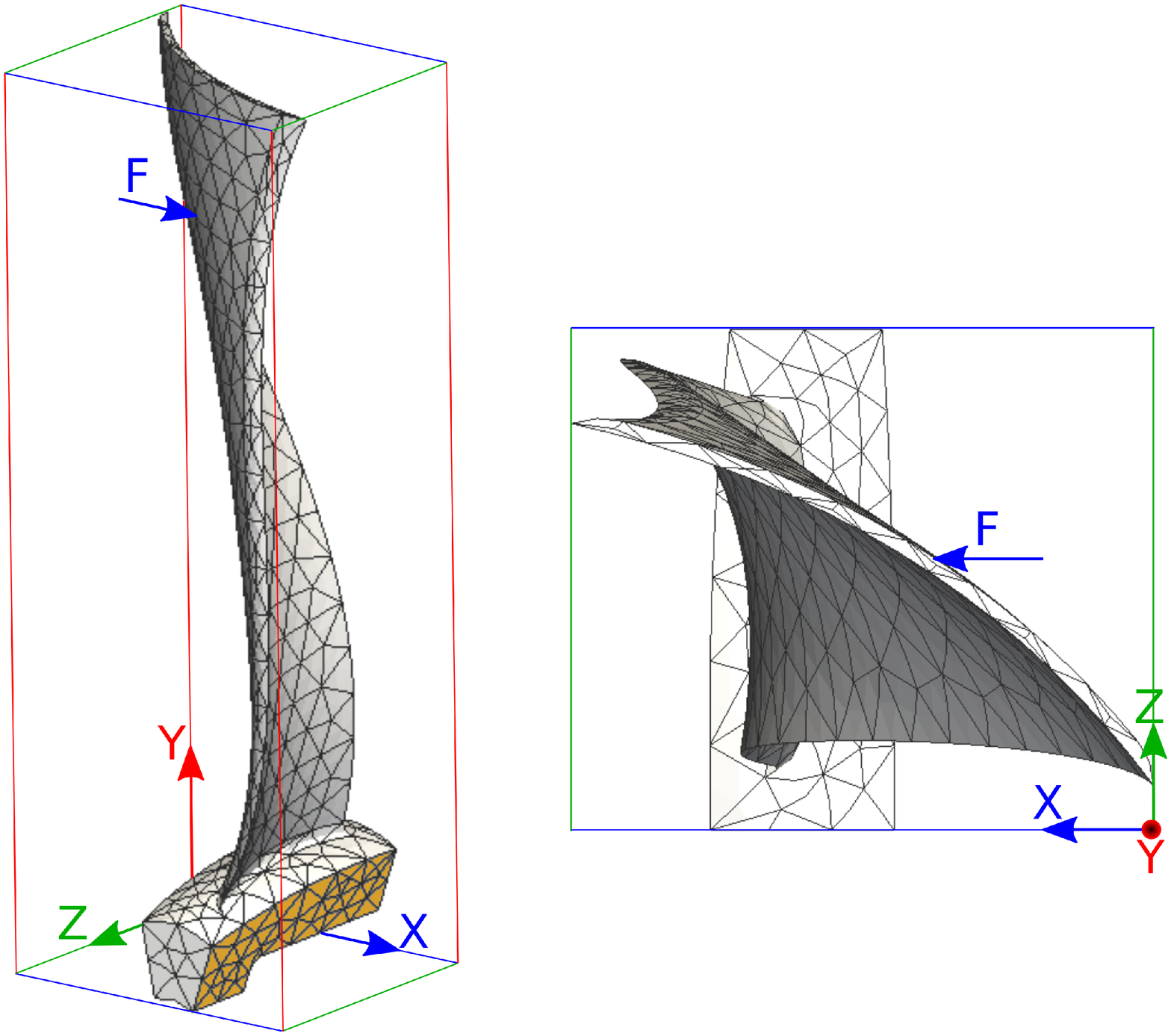}
\caption{\footnotesize Blade mesh isometric and top view.}\label{fig:blade_mesh}
\end{subfigure}
\begin{subfigure}{.58\textwidth}\centering
\includegraphics[height=4.5cm]{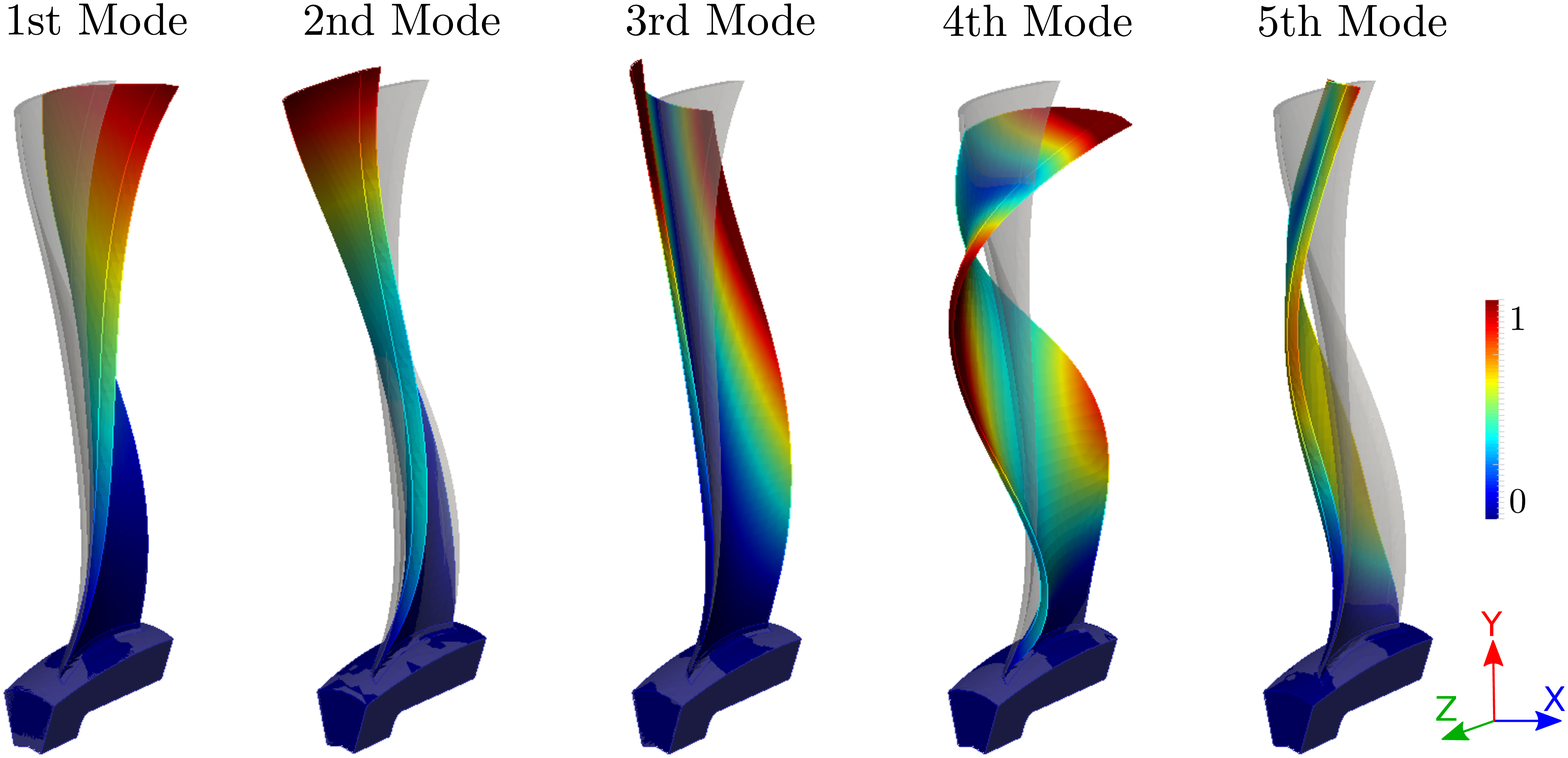}
\caption{\footnotesize Blade modes}\label{fig:blade_modes}
\end{subfigure}
\caption{Blade mesh and first five modes. In Fig.~\ref{fig:blade_mesh}, the blade mesh is shown from two different views. Clamped faces highlighted in yellow. In Fig.~\ref{fig:blade_modes}, the first five modes of the blade are reported.}
\end{figure}

The FE model of the fan blade is shown in Fig.~\ref{fig:blade_mesh}, and consists of a mesh of 2041 tetrahedrons with 10 nodes each, for a total number of 3895 nodes (11685 dofs). The blade is clamped at its root as shown in Fig.~\ref{fig:blade_mesh}. The dimension of the bounding box containing the blade and the portion of disk at its root is: $\Delta X=0.41$ m, $\Delta Y=1.14$ m, $\Delta Z=0.35$ m. The span of the blade is approximately $1.02$ m, the chord measured from trailing to leading edge is $0.48$ m and the thickness in the centre of the profile is $0.01$ m. The material properties are the following: $E=104$ GPa, $\rho=4400$ kg/m$^3$, $\nu=0.3$.
The frequency of the first five modes are reported in Table~\ref{tab:freq_blade}. The mode shapes are reported in Fig.~\ref{fig:blade_modes}.

\begin{table}[h!]\centering\small
\begin{tabular}{|l| r r r r r|}
\hline
\rule{0pt}{13pt}Mode Number  &
1&   2&   3&   4&   5\\
\hline
\rule{0pt}{13pt}Frequency (Hz)& 
14.592 & 40.571 & 49.055 &  170.36 & 197.03\\
Frequency Ratio& 
1&   2.78&   3.36&    11.67&   13.50\\
\hline
\end{tabular}
\caption{First five eigenfrequencies of the fan blade and frequency ratios with the first mode.}
\label{tab:freq_blade}
\end{table}


This example is challenging for a number of reasons. First the nonlinear behaviour of the blade is close to a cantilever beam, for which inertia nonlinearity plays an important role~\cite{Farokhi2020}. In this case, reduction methods based on static condensation are not reliable anymore since neglecting the most important phenomenon.  Second, as compared to the beam, the shape of the fan is twisted and curved, making the problem closer to a shell than to a flat structure. Consequently quadratic nonlinearities are much more pronounced and slow/fast separation is less fulfilled. For all these reasons, computing the correct type of nonlinearity (hardening/softening behaviour) of this kind of structure, still remains difficult for a ROM. As a matter of fact, no clear numerical demonstration that a reduced-order model is able to accurately compute the non-linear vibration of a Fan Blade has been proposed so far.  The existing methods published in literature for vibration of large 3D models cannot achieve a better accuracy that $10\%$ of error, in the best case, as compared to the full model~\cite{Balmaseda2020}.

\begin{figure}[h!]
\centering
\includegraphics[height=4.5cm]{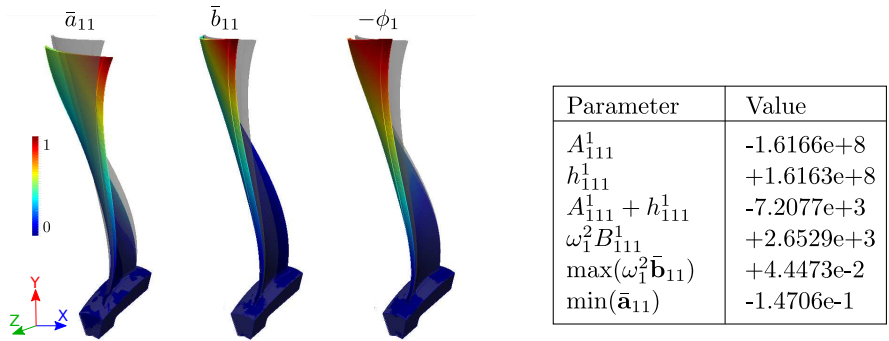}
\caption{Second-order correction vector shapes $\ag_{11}$ and $\bg_{11}$ used in the reduction procedure for the first mode of the fan blade $\phi_1$. The vector $\bg_{11}$ presents a strong component of the mode  $\phi_1$, highlighting the important quadratic self-coupling term $g^1_{11}$ due to the curvature. In the table are listed the values of the parameters needed to construct the reduced-order dynamics for the fundamental mode.}
\label{fig:tensors_blade}
\end{figure}

In Fig.~\ref{fig:tensors_blade}, the second order tensors $\ag,\bg$ relative to the first mode of the blade are reported together with the reduced dynamics parameters of the ROM for a single mode motion. The shape of the $\ag$ vector shows that a shortening motion (proportional to $\R^2$) takes place as the vibration amplitude of the first mode (proportional to $\R$) increases. This is a fairly expectable behaviour occurring in every cantilever-like structure. The correction provided by $\ag$ thus serves the purpose of relaxing the stiffening effect that one would observe when this shortening motion is erroneously locked, see {\em e.g.}~\cite{Rutzmoser,RutzThesis}. The relaxation provided by $\ag$ is what generates the $A^1_{111}$ term in the reduced dynamics of the first mode. Importantly, the absolute values of $A^1_{111}$ and $h^1_{111}$ are nearly the same and their sum is almost vanishing, meaning that the $\R^3$ term is small in the reduced dynamics. Unlike overconstrained structures, cantilever-like structures are isostatic therefore the slave modes quadratically coupled to the master are free to move, generating this particular behaviour that we have also observed in other cases. 


Since the correction brought by $A^1_{111}$ makes the term $A^1_{111}+h^1_{111}$ very small, one understands that much larger amplitudes needs to be reached in order to observe a remarkable curvature in the backbone curve. Secondly, the $B^1_{111}$ term in the reduced dynamics becomes more important and convey a significant information in order to retrieve the correct hardening/softening behaviour. 
These remarks underline why reduction methods such as static and implicit condensation cannot catch properly the correct behaviour
in this particular case of cantilever structure~\cite{KimCantilever}.  Indeed, since these methods neglect velocities and inertia effects in their development, referring to a static manifold instead of the correct invariant manifold, they can retrieve the relaxation on cubic term but will neglect the inertial effect appearing in $B^1_{111}$, linked to a $R\dot{R}^2$ monomial term. As reported in~\cite{Rutzmoser}, static modal derivatives also encounters convergence issues with cantilever structures when taking into account two master modes in the ROM.


The computation of the full-order solution also presents some challenges.
As known from nonlinear continuation analysis on such complex structures, increasing the number of harmonics can make the solution more and more difficult to achieve since more and more complex internal resonance with numerous loops appears in the backbone. This has also been the case with this fan blade. A full model computed using a harmonic balance continuation method with $5$ harmonics has been found unable to achieve moderate amplitudes of vibrations, since the computation was not able to get out of numerous loops of high-order internal resonance occurring at small amplitudes. 

For such complex structures with large modal density, it is known that an impressive number of high-order internal resonance are possible creating an intricate web of loops. Internal resonance has already been observed and studied on a reduced order model of a full 3D fan blade~\cite{DiPalma2019}. Although the dynamics observed in this paper is interesting, the selected reduced order model is too simple to validate the conclusions. The authors used indeed a reduced-order model based on two linear modes with computation of the quadratic and cubic term using the STEP method.

\begin{figure}[h!]
\centering
\includegraphics[height=6cm]{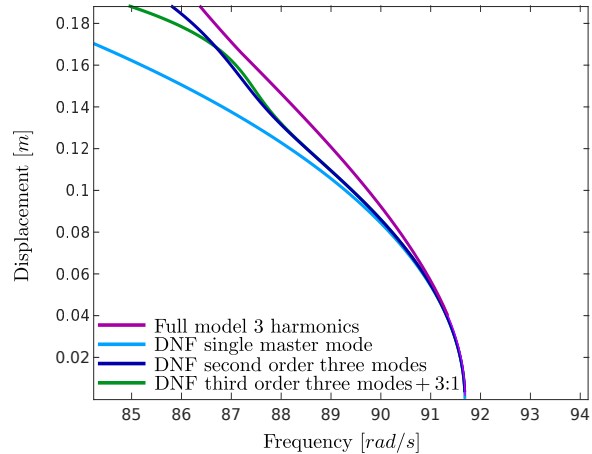}
\caption{Backbone curve of the fundamental mode of the fan blade. Amplitude of vibrations (in $m$) computed at the node where the force vector is shown, see Fig.~\ref{fig:blade_mesh}, and in the same direction $x$.}
\label{fig:all_results_blade}
\end{figure}

Also, a number of these loops occur on very short parameter range and are often not robust to {\em e.g.} addition of damping. It results that they can be interpreted as obstacle for full order simulation with continuation, that are in general not meaningful for the global observable dynamics in forced-damped case. A computation with only $3$ harmonics has been able to get rid of this limitation and attains large amplitude by filtering the higher order interactions. Hence this solution will be taken as reference, and is shown in Fig.~\ref{fig:all_results_blade}, for a vibration amplitude up to 0.18 m.
Here the reported amplitude corresponds to the displacement in the $x$ direction, at the point shown in Fig.~\ref{fig:blade_mesh}. Given the slenderness of the blade, $0.18m$ vibration amplitude at this point corresponds to $0.267m$ at the leading edge which is indeed a large magnitude with regard to usual applications. In this range, a softening behaviour is reported, in the line of usual results reported in different studies for fan blades, and showing the importance of the quadratic nonlinearity and shell-like behaviour of this structure.



Three different ROMs are built in order to test their ability to recover the backbone of the reference solution. The first ROM is composed of a single master mode. As shown in Fig.~\ref{fig:all_results_blade}, it recovers directly the correct softening behaviour, as it could have been expected from previous theoretical derivations~\cite{touze03-NNM,touze-shelltypeNL}. This good prediction is an excellent result since many reduction method encounters difficulties in retrieving such a behaviour~\cite{Rutzmoser,Balmaseda2020}. However, one can see that, when vibration amplitude at the forcing point reach values at about 0.1 m, the single NNM solution starts to depart from the reference solution, underlining that a new stiffening effect comes into play and needs to be taken into account on order to obtain a ROM prediction closer to the reference.

As shown in Table~\ref{tab:freq_blade}, modes 2 and 3 of the blade have close eigenfrequencies as compared to the first. Also, they are both not far from showing a 3:1 resonance relationship with the first eigenfrequency. Indeed, the ratio are respectively $2.78$ and $3.36$. Depending on the hardening/softening behaviour of each mode, it is likely that the exact fulfilment of 3:1 resonance might appear at larger amplitudes. Consequently, two other ROMs have been built, including these two additional modes. The first one is thus composed of master modes 1, 2 and 3, and uses second-order normal form, while the second one selects the same master modes, but uses third-order normal form with additional resonant terms corresponding to 3:1 resonance relationships. These two ROMs predicts backbone curves that are closer to the reference, showing that the addition of these supplementary modes is meaningful for achieving a correct solution on a larger range of amplitudes. They nevertheless show a slight departure from reference at large amplitude, underlining the fact that the full-order solution is again more complex, with possibly other higher-order resonances appearing.


\subsection{Frequency-response curve with damping and forcing}\label{sec:FRF}
In this last section, frequency-response functions (FRFs) are computed including damping and forcing.
The two structures of the previous sections, namely the straight beam and the fan blade, are used
for illustrative purposes. The aim is to show how on can easily build ROM including damping and forcing
from the DNF technique, using the guidelines provided in section~\ref{sec:dampforc}.
First the fan blade is used to highlight the efficiency of the damping model proposed, by comparing the
results when using only either mass-proportional or stiffness-proportional damping. Then the beam is considered in order to show the behaviour of the FRF when approaching the 3:1 internal resonance, in the case of the second NNM.

\subsubsection{Fan Blade}

The frequency response curve of the fan blade investigated in section~\ref{sec:blade} are computed and comparison between a full-order simulation and  reduced-order models composed of a single master mode using DNF with damping and forcing, is produced. The forcing is harmonic with forcing frequency $\Omega$ and magnitude $F$. It is aligned with the direction $x$ and located at the point shown in Fig.~\ref{fig:blade_mesh}, a few centimetres below the tip. To build the ROM, the guidelines provided in section~\ref{sec:dampforc} are applied. A single master coordinate corresponding to mode 1 is selected. A modal force corresponding to modal projection of the external forcing is added at the right-hand side of the dynamics. A linear damping term, directly deduced from the Rayleigh damping coefficients, is taken into account, as well as the additional nonlinear damping term involving the coefficient $C^1_{111}$, and aggregating the contributions of the damping factors of the slave modes. In the remainder, comparisons will be drawn out by considering this additional term or not.

In the first case under study, only mass-proportional damping is considered. The amplitude of the forcing is set at $F=30N$, and the Rayleigh coefficients are such that $\zeta_M=3E-1$ and $\zeta_K=0$. In the case of only mass-proportional damping, the modal damping factors reads $\xi_p=\zeta_M/2\omega_p$, where $\xi_p$ is the nondimensional factor appearing when expressing the linear damping term as $2\xi_p\omega_p\dot{X}_p$. Consequently this damping factor decreases with the frequency, which is not a correct representation of real structures where losses generally increases with frequency. In this specific case, the slave modes are thus less and less damped and their influence on the master low-frequency mode that is directly excited is awaited to be negligible.

\begin{figure}[h!]
\centering
\includegraphics[height=6cm]{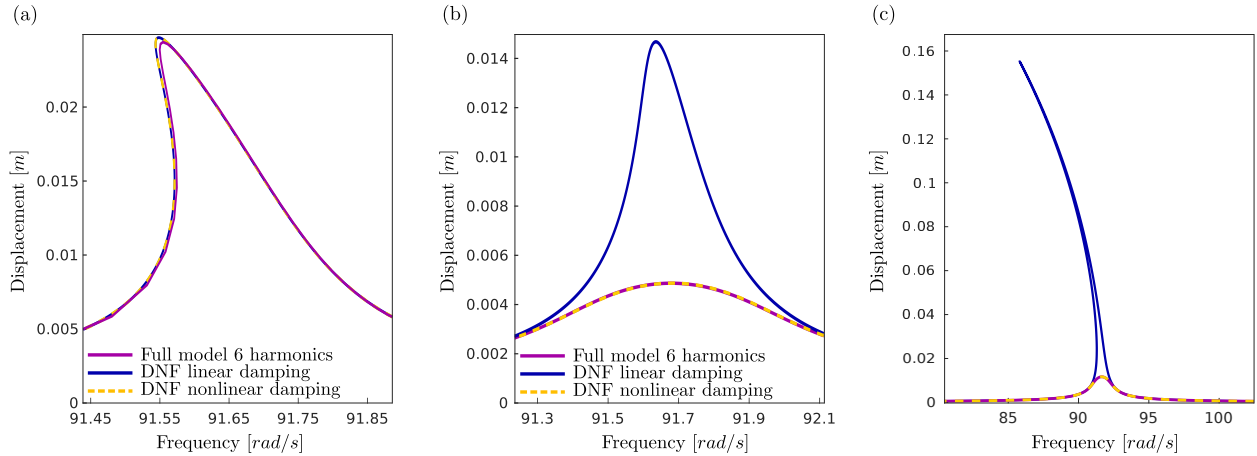}
\caption{Frequency response functions of the fan blade subjected to harmonic forcing in the vicinity of the first bending eigenfrequency. (a) Only mass-proportional Rayleigh damping is taken into account, with $\zeta_M=3E-1$ and $\zeta_K=0$. Amplitude of the forcing $F=30N$. Comparison between full-order model (purple), and ROM composed of a single master mode computed with direct normal form, either with only linear damping term (DNF linear damping, blue curve), or by taking into account the nonlinear aggregated damping term (DNF nonlinear damping, yellow dashed curve). (b) Only stiffness-proportional damping with $\zeta_K=2E-5$ and $\zeta_M=0$. Amplitude of forcing $F=2N$. (c)  Only stiffness-proportional damping with $\zeta_K=1E-5$, $\zeta_M=0$, amplitude of forcing $F=10N$.}
\label{fig:fanbladedampingF}
\end{figure}

Fig.~\ref{fig:fanbladedampingF}(a) shows the FRFs obtained and compares the full-order solution to two different ROMs, one where only the linear damping term $\zeta_M\dot{R}_1$ is taken into account in the ROM (DNF linear damping, blue curve), and another one where the aggregated nonlinear damping term is added (DNF nonlinear damping, dashed yellow curve). In this case the two ROMs reproduce very well the solution of the full-order model, and the difference between the two is negligible, meaning that the aggregated $C^1_{111}R_1^2\dot{R}_1$ is very small and can be easily neglected. This is the direct consequence of the choice of only mass-proportional damping and the fact that modal damping ratios are decreasing with frequencies.

Fig.~\ref{fig:fanbladedampingF}(b) and (c) shows two different cases where only stiffness-proportional damping is selected. This case is much more interesting since it corresponds to a more physically relevant situation. Also, since the modal damping factors are linearly increasing in this case, this means that the damping factors of the slave modes are more and more important, so that neglecting their effect in a ROM would lead to discrepancies. Fig.~\ref{fig:fanbladedampingF}(b) shows a case of an important stiffness-proportional damping with $\zeta_K=2E-5$ and $\zeta_M=0$. The forcing is set to $F=2N$. One can now observe a very important difference between the FRFs computed by the two ROMs. Neglecting the aggregated nonlinear damping term leads to important overestimation of the amplitude of the solution in the resonant region. On the other hand, the ROM with the nonlinear damping exactly reproduces the amplitude of the full model, and enforces the structure to vibrate in an almost linear regime as it should be.

Fig.~\ref{fig:fanbladedampingF}(c) shows a case with smaller value of damping,  $\zeta_K=1E-5$, $\zeta_M=0$, and larger amplitude of forcing: $F=10N$. Even in this case the behaviour of the full-order model is almost linear with a small bump close to linear resonance. Interestingly, the prediction given by taking into account only the linear damping term is completely wrong and gives amplitudes that are huge as compared to the reference. This clearly shows that neglecting the damping of the slave modes can lead to erroneous predictions. On the other hand, the nonlinear damping term exactly recovers the correct amplitude and is able to perfectly match the reference solution. These examples shows that even in a system with complex geometry and 3D elements, the proposed ROMs are able to compute with great accuracy the FRF with minimal models.
\subsubsection{Straight beam with internal resonance}

In this last example, FRFs of the straight clamped-clamped beam considered in section~\ref{sec:beam} are considered. The idea is to show how the method behaves in case of internal resonance where there is the need to take into account more than one master coordinate in the ROM. For that purpose, a harmonic forcing in the vicinity of the second eigenfrequency is considered. Two different values of forcing are selected. The first case is that of a small amplitude of forcing, such that the FRF does not enter in the loop of the 3:1 resonance. In this case, only one master mode is awaited to produce a correct prediction. A second case of larger amplitude is then selected so that the FRF enters the resonance loop where 3:1 with mode 4 is excited. In this case, ROMs built from two master coordinates will be selected. In each case the forcing is located at $z=0.275$ from the clamp as in Fig.~\ref{fig:beam_mesh}, and excites the beam in the $x$ direction. Finally, only the most interesting case of only stiffness-proportional damping is considered, since it is closer to real situations and give rise to more important differences between models.

\begin{figure}[h!]
\centering
\includegraphics[height=6cm]{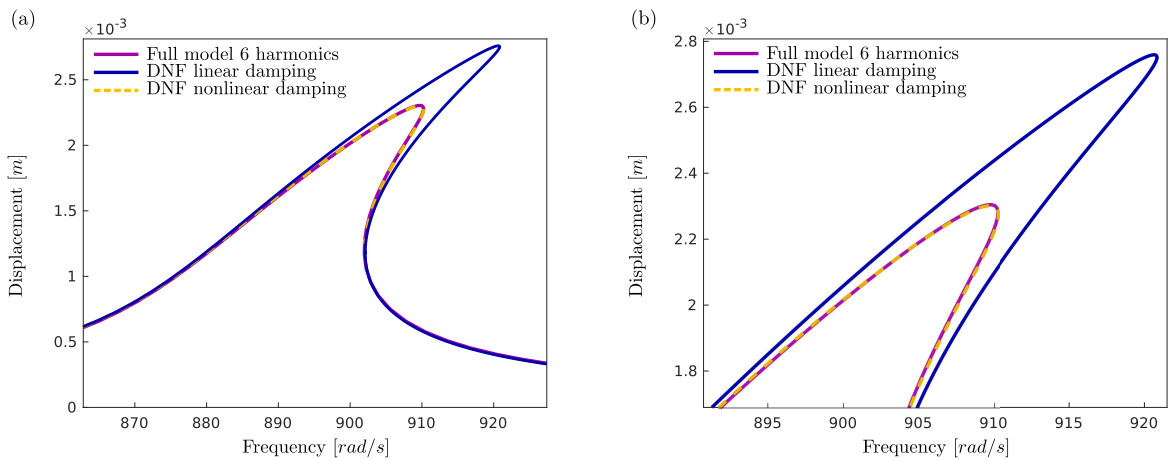}
\caption{(a) Frequency response functions of the clamped-clamped beam subjected to harmonic forcing in the vicinity of the second eigenfrequency. Only stiffness-proportional damping is considered with $\zeta_K=13E-6$ and $\zeta_M=0$. Amplitude of the forcing $F=10N$. Full-order solution is compared to ROMs obtained with a single master coordinate (mode 2), with only linear damping term (blue curve) and with additional nonlinear damping term (dashed yellow curve). (b) close-up view near the maximum amplitude.}
\label{fig:damped_beam}
\end{figure}

Fig.~\ref{fig:damped_beam} shows the results obtained when selecting $\zeta_K=13E-6$ and $F=10N$, such that the FRF is below the internal resonance loop. The reference solution is obtained by taking into account 6 harmonics in the expansion. Two different ROMs are contrasted, both containing a single master coordinate corresponding to mode 2, and one with only the linear damping term while the second one takes into account the aggregated nonlinear damping term. As in the previous case, the difference between the two ROMs are important and taking into account the nonlinear damping term is very important in order to correctly predict the maximum amplitude of the FRF.

\begin{figure}[h!]
\centering
\includegraphics[height=6cm]{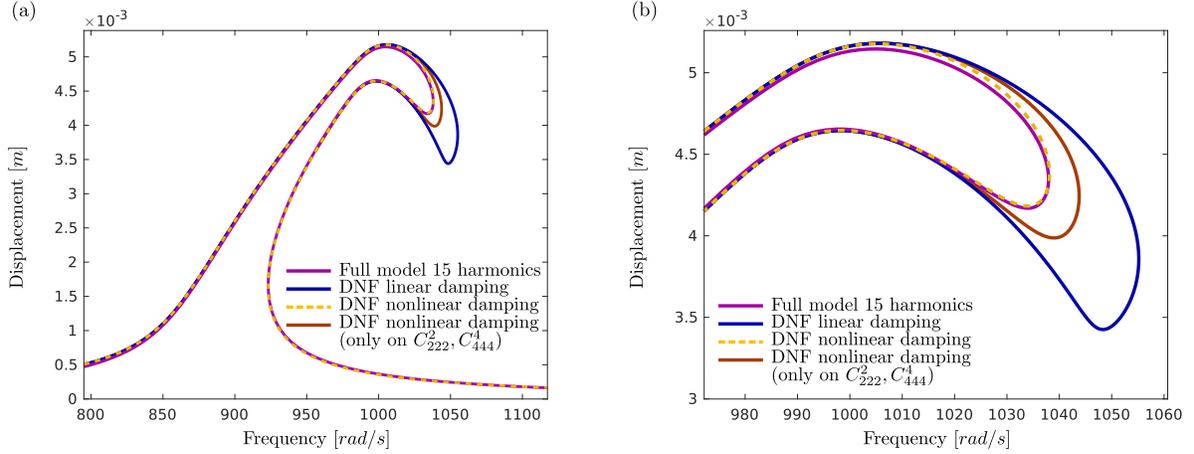}
\caption{(a) Frequency response functions of the clamped-clamped beam subjected to harmonic forcing in the vicinity of the second eigenfrequency. Stiffness-proportional damping is considered with $\zeta_K=3E-7$ and $\zeta_M=0$. Amplitude of the forcing $F=30N$. Full-order solution is compared to different ROMs obtained with a two master coordinates corresponding to modes 2 and 4. Reference solution obtained with 15 harmonics (purple). All the ROMs are obtained with two modes and  second-order normal form. Blue curve: only linear damping term considered. Yellow dashed curve: full consideration of nonlinear damping including cross-coupling terms. Brown curve: nonlinear damping neglecting the cross-couplings. (b) close-up view in the 3:1 resonance loop.}
\label{fig:damped_beam_IR}
\end{figure}

Fig.~\ref{fig:damped_beam_IR} now considers the case of a smaller damping coefficient $\zeta_K=3E-7$ and a larger amplitude of forcing $F=30N$. The reference solution now enters the loop so that the 3:1 internal resonance with mode 4 is activated. Consequently, the reference solution has been computed with a larger number of harmonics (15), in order to achieve convergence. Three different ROMs are compared, all of them being built with two master coordinates corresponding to modes 2 and 4, and second-order normal form. The first ROM contains only the linear viscous damping term. Since the underlying conservative system with second-order normal form is able to perfectly well reproduce the resonance loop of the 3:1 resonance, this solution also shows this typical feature, highlighting that the dynamics is more complex and involves more than one mode. However, the damping is not important enough, so that a slight departure is observed from the reference solution. A second ROM is built by adding nonlinear damping terms, following Eq.~\eqref{eq:ROM_SO_damped}. However a simplification is introduced. In the summation terms involving $C^p_{ijk}$ terms, only the self-nonlinear damping terms $C^p_{ppp}$ are considered. This means that only two terms have been added to the reduced dynamics: $C^2_{222}R_2^2\dot{R}_2$ for the equation of mode 2 (first master coordinate) and $C^4_{444}R_4^2\dot{R}_4$ for the equation of mode 4 (second master coordinate). The cross-coupling nonlinear damping terms involving $C^p_{ijk}$ with different indexes have been discarded. In this case, one can observe a better behaviour as compared to the linear case, meaning that the two added terms convey important information on the damping of the slave modes, that cannot be neglected. However the result is still a bit different from the reference solution. 

Finally a ROM with two modes and considering all the nonlinear damping terms including the cross-couplings one, is reproduced in Fig.~\ref{fig:damped_beam_IR}, and shows an almost perfect comparison with the reference. This example shows that these cross-coupling terms also convey an important information, needed in order to reproduce with great accuracy the FRF.

\section{Conclusion}
A nonlinear mapping for the derivation of reduced-order models for geometrically nonlinear structures discretised by the FE method, has been presented. It relies on the normal form theory and presents a firm theoretical background in order to cope with nonlinear dynamics phenomena such as internal resonance. The method allows for a direct computation of the normal form, from a FE discretisation. Computational details are provided and a full algorithm has been presented with related discussion on the non-intrusiveness of the method and coding advice for practical implementation. In essence, the technique allows one to go from the physical space (nodes of the FE structure given by the original mesh) to an invariant-based span of the phase space, with a third-order approximation of the invariant manifolds (nonlinear normal modes) of the structure. It is a simulation-free method that directly computes the third-order approximation of the reduced dynamics, with no restriction on the number of selected master modes.

The main developments have been proposed for a correct estimation of the nonlinear force of the ROM, {\em i.e.} in a conservative framework. Two declinations of the method have been proposed: a second-order normal form and a third-order mapping, with, in each case, the associated reduced dynamics. The second-order normal form proposes a quadratic mapping approximating the underlying invariant manifold up to the second-order only. Consequently, the reduced dynamics (up to order three) contains all possible third- and higher-order internal resonances embedded, and has no quadratic nonlinearity if no second-order internal resonance are present. The third-order normal form distinguishes the trivial and non-trivial resonant terms. The main improvement consists in offering a better approximation of the invariant manifolds, and a more simplified reduced dynamics with only trivially resonant monomial cubic terms included. On the other hand, the reduced dynamics contains less terms since the higher-orders have been neglected. Consequently the dynamical solutions of the second-order DNF can produced more accurate results as highlighted in the examples.

A methodology has been proposed in order to take into account external forcing and damping in the ROM so as to produce frequency-response functions (FRF). For the damping, the proposed technique relies on a first-order approximation of the general result derived in~\cite{TOUZE:JSV:2006}. It has been rewritten for Rayleigh damping law, the most commonly retained assumption to model losses in FE context. The main interest relies in proposing an aggregated damping force on the reduced dynamics, taking into account the losses of all the slave (neglected) modes. In the present version, the general formulas for taking properly into account the damping have been given only for the second-order DNF.

The method has been applied to a clamped-clamped beam and a FE model of a fan blade. Backbone curves have been extracted and discussed. In particular, as known from previous studies on normal form for model order reduction, the method predicts directly the correct hardening/softening behaviour. Also, in the beam case, it has been shown that the ROM can recover internal resonance loops due to the fulfilment of an exact commensurability of nonlinear frequencies. FRF have been computed, underlining the importance of the added, aggregated damping, as well as the very good match between full and reduced models.

Future work will apply the methodology to different structures having also internal resonance between eigenfrequencies  of the system, in order to test the predictions of the ROM with an increasing number of master modes. Improvements of the method can also be investigated. For example higher-order reduced dynamics can be computed by pushing the normal form up to order five for better accuracy. Taking into account the full damped version of the normal form as developed in~\cite{TOUZE:JSV:2006} could also be helpful. Finally, different internal forces could also be taken into account by enlarging the application field to {\em e.g.} gyroscopic forces, electrostatic forces (for MEMs applications), non conservative follower forces (for fluid-structure interaction problems) or piezoelectric couplings.

\section*{Conflict of interest}
The authors declare that they have no conflict of interest.
\section*{Codes availability statement}
The codes written to run most of the simulations presented in this paper can be available upon simple request to the authors. Note that for all the simulations the open-source software \texttt{code\_aster} has been used. Future  developments of this code should include the reduction method proposed in this paper as a meta-command.
\section*{Acknowledgements}
The author A. Vizzaccaro is thankful to Rolls-Royce plc for the financial support. The author Y. Shen wishes to thank China Scholarship Council (No.201806230253). The author L. Salles is thankful to Rolls-Royce plc and the EPSRC for the support under the Prosperity Partnership Grant ”Cornerstone: Mechanical Engineering Science to Enable Aero Propulsion Futures”, Grant Ref: EP/R004951/1. The author J. Blaho\v{s} thank the European Union’s Horizon 2020 Framework Programme research and innovation programme under the Marie Sklodowska-Curie agreement No 721865.
\bibliographystyle{unsrt}
\bibliography{biblio}

\appendix

\section{Derivation of second-order tensor}\label{app:atobara}

In this appendix, we start by recalling the general formulas needed to compute the second-order tensors from the normal form in modal basis, namely $\at$, $\bt$, $\ct$. These have been first derived in~\cite{touzeLMA,touze03-NNM}, they are here expressed in a different setting allowing for better comparison with the tensors obtained with the direct method. 

The derivation of the second order tensors in modal basis, is obtained by substituting the nonlinear mapping equations for $\x,\y$ into the equations of motion for $\dot{\x},\dot{\y}$, by replacing each $\dot{\R}_r$ (and $\dot{S}_r$) term with the first order assumption $\dot{\R}_r=\S_r$ (and $\dot{S}_r=-\omega_r^2\R_r$), and finally by balancing the second order terms. This process will lead to four equations for each couple of indexes $i,j$, one for the terms multiplying $\R_i \R_j$, one for $\S_i \S_j$, one for $\R_i \S_j$, and one for $\S_i \R_j$. The equations obtained permit to derive $\at_{ij},\bt_{ij},\ct_{ij},\ct_{ji}$; the system reads:
\begin{equation}
\begin{cases}
2\at_{ij}-2\omega_i^2\bt_{ij}-\ct_{ij}=\zervec,\\
2\at_{ij}-2\omega_j^2\bt_{ij}-\ct_{ji}=\zervec,\\
-2\Ot\at_{ij}+\omega_j^2\ct_{ij}+\omega_i^2\ct_{ji}=\g_{ij},\\
2\Ot\bt_{ij}+\ct_{ij}+\ct_{ji}=\zervec.
\end{cases}
\end{equation}
The solution of this system of equations in the unknown tensors can be written as:

\begin{subequations}\begin{align}
\mathbf{D}_{ij}\at_{ij}=&
((+\omega_i^2+\omega_j^2)\I-\Ot)
 \g_{ij},
\\
\mathbf{D}_{ij}\bt_{ij}=&
2\, \g_{ij},
\\
\mathbf{D}_{ij}\ct_{ij}=&
((-\omega_i^2+\omega_j^2)\I-\Ot)
2\,\g_{ij},
\\
\mathbf{D}_{ij}\ct_{ji}=&
((+\omega_i^2-\omega_j^2)\I-\Ot)
2\,\g_{ij},
\end{align}\label{eq:second_order_tensors_modal}\end{subequations}

where the left-hand side term $\mathbf{D}_{ij}$ is the denominator of the coefficients. It makes appear the second order internal resonance relationship (small denominator problem), and reads:
\begin{equation}
\mathbf{D}_{ij}=
((+\omega_i+\omega_j)^2\I-\Ot)
((-\omega_i+\omega_j)^2\I-\Ot).
\end{equation}
If no second order internal resonance occurs, in Eqs.~\eqref{eq:second_order_tensors_modal} the denominator matrix can be inverted and the expression for the tensors made explicit.
With these expressions, one is then able to compute all the coefficients needed for the modal normal form.

In order to shed more light on the close relationship between the coefficients of the modal and the direct normal form, let us show on a specific term $\modal{a}^s_{ij}$ of the second order tensor $\at_{ij}$ how one can go from one form to another. The explicit expression for $\modal{a}^s_{ij}$ can be deduced from the previous equation or found in~\cite{touze03-NNM}. It reads:
\begin{equation}
\modal{a}^s_{ij}=
\dfrac{
((+\omega_i^2+\omega_j^2)-\omega_s^2)
}{
((+\omega_i+\omega_j)^2-\omega_s^2)
((-\omega_i+\omega_j)^2-\omega_s^2)
}
\modal{g}^s_{ij},
\label{eq:at_non_split}
\end{equation}

Let us first  rewrite this formula so that the denominator, which is of order four in $\omega$, is split into two factors of order two, reminding that this denominator has a physical meaning. Indeed, the two factors represents the two possible second-order internal resonances between the modes of index $i,j,s$. One obtains:
\begin{equation}
\modal{a}^s_{ij}=
\left(
\dfrac{1}{(+\omega_i+\omega_j)^2-\omega_s^2}
+
\dfrac{1}{(-\omega_i+\omega_j)^2-\omega_s^2}
\right)
\dfrac{\modal{g}^s_{ij}}{2}.
\label{eq:at_split}
\end{equation}
This can be rewritten in a more compact form for the vector $\at_{ij}$ for all $s$:
\begin{equation}
\at_{ij}=
\dfrac{1}{2}((+\omega_i+\omega_j)^2\I-\Ot)^{-1} \g_{ij}
\;+\;
\dfrac{1}{2}((-\omega_i+\omega_j)^2\I-\Ot)^{-1} \g_{ij}.
\end{equation}
It is now possible to express $\at$ in terms of $\K$ and $\M$ by using Eqs.~\eqref{eq:diagonalisation} stating $\V^T\M\V=\I$, and $\V^T\K\V=\Ot$, so that:
\begin{equation}
\at_{ij}=
\dfrac{1}{2}((+\omega_i+\omega_j)^2\V^\text{T}\M\V-\V^\text{T}\K\V)^{-1} \g_{ij}
\;+\;
\dfrac{1}{2}((-\omega_i+\omega_j)^2\V^\text{T}\M\V-\V^\text{T}\K\V)^{-1} \g_{ij}.
\label{eq:a_with_V}
\end{equation}
Each factor in brackets can be rewritten as:
\begin{equation}
\left(\V^\text{T}\, (\sigma_{ij}^2\M-\K)\, \V\right)^{-1}
=\V^{-1}\, (\sigma_{ij}^2\M-\K)^{-1}\, \V^{-\text{T}}
\end{equation}
where for the sake of readability the term $\sigma_{ij}$ replaced the summation of eigenvalues $(\pm\omega_i+\omega_j)$. One can see that the full eigenvector matrix appears twice in each factor. Consequently a simplification arises by premultiplying Eq.~\eqref{eq:a_with_V} by $\V$, which also leads to make appear  $\ag_{ij}$ since  $\ag_{ij}=\V \at_{ij} $. Also, by noticing that $\V^{-\text{T}} \g_{ij} = \G (\phit_i, \phit_j)$, the post-multiplied term can be simplified, so that a final expression involving only terms from the physical basis is obtained:
\begin{equation}
\ag_{ij}=
\dfrac{1}{2}((+\omega_i+\omega_j)^2\M-\K)^{-1} \G(\phit_i,\phit_j)
\;+\;
\dfrac{1}{2}((-\omega_i+\omega_j)^2\M-\K)^{-1} \G(\phit_i,\phit_j).
\label{eq:a_physapp}
\end{equation}
Lastly, by defining the two vectors $\Zd_{ij},\Zs_{ij}$ the expressions given in text can be finally retrieved. Moreover, by applying a similar procedure to the other tensors $\bt,\ct$, one can show that the four tensors can all be expressed as linear combinations of $\Zd_{ij},\Zs_{ij}$, hence only two operations are needed to derive four tensors.

Another possible procedure to express the tensors in physical coordinates, is now described in order to show why the splitting operation done in Eq.~\eqref{eq:at_split} has permitted to drastically reduce the computational cost. In fact, to derive the expression of the tensors in physical basis, the splitting operation is not strictly needed and another expression for them can be found starting from Eq.~\eqref{eq:at_non_split}. From Eq.~\eqref{eq:at_non_split}, the expression for $\at_{ij}$ in compact form reads:
\begin{equation}
\at_{ij}=
((-\omega_i+\omega_j)^2\I-\Ot)^{-1}
((+\omega_i+\omega_j)^2\I-\Ot)^{-1}
((+\omega_i^2+\omega_j^2)\I-\Ot)
 \g_{ij}.
 \label{eq:at_modal_coord}
\end{equation}
Again, its equivalent in physical coordinates is the tensor $\ag_{ij} = \V\at_{ij}$, and the equivalent of the forcing term in physical coordinates is $\G(\phit_i,\phit_j)=\V^{-T}\g_{ij}$. By introducing these two equivalences into Eq.~\eqref{eq:at_modal_coord} one obtains:
\begin{equation}
\ag_{ij}=
\V
((-\omega_i+\omega_j)^2\I-\Ot)^{-1}
((+\omega_i+\omega_j)^2\I-\Ot)^{-1}
((+\omega_i^2+\omega_j^2)\I-\Ot)
 \V^{T}\G(\phit_i,\phit_j).
 \label{eq:at_phys_coord_1}
\end{equation}
This equation still presents two matrices in modal coordinates: the full eigenvectors matrix $\V$ and the full eigenvalues matrix $\Ot$. To eliminate these two matrices and make $\K$ and $\M$ appear, one last operation is needed. Introducing the matrix:
\begin{equation}
\Og = \M^{-1}\K,
\end{equation}
one has that the relationship between $\Og$ and $\Ot$ is:
\begin{equation}
\Ot = \V^{-1}\Og\V.
\end{equation}
Finally, by substituting into Eq.~\eqref{eq:at_phys_coord_1}, the last expression for $\Ot$ and simply $\V^{-1}\V$ for $\I$, the direct equation for $\at_{ij}$ can be obtained:
\begin{equation}
\ag_{ij}=
((-\omega_i+\omega_j)^2\I-\Og)^{-1}
((+\omega_i+\omega_j)^2\I-\Og)^{-1}
((+\omega_i^2+\omega_j^2)\I-\Og)
\M^{-1}\G(\phit_i,\phit_j).
 \label{eq:at_phys_coord_2}
\end{equation}
This expression of $\at$ has been given in \cite{vizzaENOC} for the case of  equal indexes $ii$. Interestingly,   it bears a strong resemblance with the expression given in \cite{VERASZTO} where the ROM is built using a direct approach based on spectral submanifold (SSM).

Comparing this expression for $\at$ with the one obtained after the splitting operation (Eq.~\eqref{eq:a_physapp}), and bearing in mind that they are completely equivalent, one sees that from a computational point of view, Eq.~\eqref{eq:at_phys_coord_2} is much more expensive. In Eq.~\eqref{eq:a_physapp}, two linear systems have to be solved to find $\Zd$ and $\Zs$ and once they are found, the four tensors $\at,\bt,\ct$ can be obtained. In Eq.~\eqref{eq:at_phys_coord_2}, three linear systems have to be solved, one for each $-1$ appearing in the formula, and these three linear systems have to be solved for each tensor, making the number of linear systems to be solved equal to twelve. However, the most computational expensive operation is not the solution of a linear system but the inversion of the matrix $\M$ needed to build the matrix $\Og$. This cost of this operation is equivalent to the cost of solving a linear system for each degree of freedom of the structure, making procedure infeasible for any engineering application. For this reason, the splitting operation that brought to Eq.~\eqref{eq:a_physapp}, although not strictly necessary from a theoretical point of view, is instead crucial from a computational one.
\section{Derivation of third-order tensor}\label{app:rtobarr}

In this appendix, a similar approach to that of Appendix~\ref{app:atobara} is used to generate the expressions of the third order tensors $\rt,\ut,\mt,\nt$. We start by recalling the general expressions from the modal approach already developed in~\cite{touzeLMA,touze03-NNM} .

In absence of second order internal resonances, no second order terms are present in the reduced dynamics equations and the first order assumptions ($\dot{\R}_r=\dot{S}_r$ and $\dot{S}_r=-\omega_r^2\R_r$), can still be used in the derivation of the third order tensors. 
The system resulting from the balance of each third order term for a given triplet of indexes $i,j,k$ reads:
\begin{equation}
\begin{cases}
3 \mt_{ijk} - \ut_{ijk} - \ut_{jki} - \ut_{kij} = \zervec,\\
3 \rt_{ijk} - \nt_{kij} - \omega_i^2\ut_{jki}  - \omega_j^2\ut_{ijk}  = \zervec,\\
3 \rt_{ijk} - \nt_{jki} - \omega_i^2\ut_{kij}  - \omega_k^2\ut_{ijk}  = \zervec,\\
3 \rt_{ijk} - \nt_{ijk} - \omega_j^2\ut_{kij}  - \omega_k^2\ut_{jki}  = \zervec,\\
\omega_i^2\nt_{ijk}  + \omega_j^2\nt_{jki}  + \omega_k^2\nt_{kij}  - 3 \Ot \rt_{ijk}  
= 3 \h_{ijk}+\At_{ijk}+\At_{jki}+\At_{kij},\\
-\nt_{jki} - \nt_{kij} + 3 \omega_i^2\mt_{ijk}  - \Ot\ut_{ijk}  = \Bt_{ijk},\\
-\nt_{ijk} - \nt_{kij} + 3 \omega_j^2\mt_{ijk}  - \Ot\ut_{jki}  = \Bt_{jki},\\
-\nt_{ijk} - \nt_{jki} + 3 \omega_k^2\mt_{ijk}  - \Ot\ut_{kij}  = \Bt_{kij}.
\end{cases}
\end{equation}
The expressions resulting from the solution of this system are much more lengthy than those of second order tensors but they possess a similar structure. For instance, the expression for the tensor $\rt$ is in the following form:
\begin{equation}
\Pol{D}{ijk}{8}\rt_{ijk}=
\;
 \Pol{Q}{ijk}{6}
 \omega_j \omega_k\Bt_{ijk}
 \;
+\;
 \Pol{Q}{jki}{6}
 \omega_k \omega_i\Bt_{jki}
 \;
+\;
 \Pol{Q}{kij}{6}
 \omega_i \omega_j\Bt_{kij}
 \;
+\;
\Pol{P}{ijk}{6}
(\At_{ijk}+\At_{jki}+\At_{kij}+3\h_{ijk}).
\label{eq:r_modal_pol}
\end{equation}
The \textit{numerator} matrices $\Pol{Q}{}{6}$ and $\Pol{P}{}{6}$ that multiplies the forcing terms $\At$ and $\Bt$ are in the form:
\begin{align}
\Pol{P}{ijk}{6}
=&
\pol{p}{ijk}{6}\I+\pol{p}{ijk}{4}\Ot+\pol{p}{ijk}{2}(\Ot)^2 + (\Ot)^3,
\\
\Pol{Q}{ijk}{6}
=&
\pol{q}{ijk}{6}\I+\pol{q}{ijk}{4}\Ot+\pol{q}{ijk}{2}(\Ot)^2 + (\Ot)^3,
\end{align}
with $\pol{p}{ijk}{\times}$ and $\pol{q}{ijk}{\times}$, both $\times$-order polynomials in the sole parameters $\omega_i,\omega_j,\omega_k$. The numerator matrices varies for the different third order tensors. Conversely, the denominator matrix is the same for each tensor and reads:
\begin{equation}
\Pol{D}{ijk}{8}=
((+\omega_i+\omega_j+\omega_k)^2\I-\Ot)
((-\omega_i+\omega_j+\omega_k)^2\I-\Ot)
((+\omega_i-\omega_j+\omega_k)^2\I-\Ot)
((+\omega_i+\omega_j-\omega_k)^2\I-\Ot).
\end{equation}
Once again, the denominator matrix carries a physical meaning in that every possible third order internal resonance between the frequency appears in it.

Following a similar approach than that on second order tensors, it is convenient to rewrite the expressions for the third order tensors in such a way that the denominator is split. Taking a combination of indexes $ijk$ with no internal resonance between their eigenvalues,
we can retrieve the basic vectors that represents the equivalent of $\Zs$ and $\Zd$ for the third order tensors in modal basis.
If one defines:
\begin{subequations}
\begin{align}
\Zat_{ijk}=((+\omega_i+\omega_j+\omega_k)^2\I-\Ot)^{-1}
\left(
\At_{ijk}+\At_{jki}+\At_{kij}+3\,\h_{ijk}
-\omega_j\omega_k\Bt_{ijk}-\omega_k\omega_i\Bt_{jki}-\omega_i\omega_j\Bt_{kij}
\right)
\\
\Zit_{ijk}=((-\omega_i+\omega_j+\omega_k)^2\I-\Ot)^{-1}
\left(
\At_{ijk}+\At_{jki}+\At_{kij}+3\,\h_{ijk}
-\omega_j\omega_k\Bt_{ijk}+\omega_k\omega_i\Bt_{jki}+\omega_i\omega_j\Bt_{kij}
\right)
\\
\Zjt_{ijk}=((+\omega_i-\omega_j+\omega_k)^2\I-\Ot)^{-1}
\left(
\At_{ijk}+\At_{jki}+\At_{kij}+3\,\h_{ijk}
+\omega_j\omega_k\Bt_{ijk}-\omega_k\omega_i\Bt_{jki}+\omega_i\omega_j\Bt_{kij}
\right)
\\
\Zkt_{ijk}=((+\omega_i+\omega_j-\omega_k)^2\I-\Ot)^{-1}
\left(
\At_{ijk}+\At_{jki}+\At_{kij}+3\,\h_{ijk}
+\omega_j\omega_k\Bt_{ijk}+\omega_k\omega_i\Bt_{jki}-\omega_i\omega_j\Bt_{kij}
\right)
\end{align}
\label{eq:z_modal}
\end{subequations}
it is possible to demonstrate that the third order tensors of modal normal form are equivalently expressed as a linear combination of the above defined vectors and their expressions read:
\begin{subequations}
\begin{align}
&\rt_{ijk}=\dfrac{\Zat_{ijk}+\Zit_{ijk}+\Zjt_{ijk}+\Zkt_{ijk}
}{12}
\\
&\ut_{ijk}=\dfrac{-\Zat_{ijk}-\Zit_{ijk}+\Zjt_{ijk}+\Zkt_{ijk} 
}{4\omega_j\omega_k}
\\
&\mt_{ijk}=\dfrac{
-(+\omega_i+\omega_j+\omega_k)\Zat_{ijk}
+(-\omega_i+\omega_j+\omega_k)\Zit_{ijk}
+(+\omega_i-\omega_j+\omega_k)\Zjt_{ijk}
+(+\omega_i+\omega_j-\omega_k)\Zkt_{ijk} 
}{12\omega_i\omega_j\omega_k}
\\
&\nt_{ijk}=\dfrac{
+(+\omega_i+\omega_j+\omega_k)\Zat_{ijk}
-(-\omega_i+\omega_j+\omega_k)\Zit_{ijk}
+(+\omega_i-\omega_j+\omega_k)\Zjt_{ijk}
+(+\omega_i+\omega_j-\omega_k)\Zkt_{ijk}
}{4\omega_i}
\end{align}
\end{subequations}

It is then straightforward to express Eqs.~\eqref{eq:z_modal} in physical basis by using the same strategy that we applied to the second order tensors. Also in this case, there is only one linear system to solve to find the four independent tensors relative to the triplet $i,j,k$ and from them all the eight third order tensors $\rg,\ug,\mg,\ng$ can be derived. As a final remark, we want to point out that an expression in physical coordinates of Eq.~\eqref{eq:r_modal_pol} could have been also derived, by replacing the matrix $\Ot$ with its equivalent $\V^{-1}\Og\V$ but in that case the computational cost of building all the tensors $\rg,\ug,\mg,\ng$ would have been even higher  than that of the second order one.

\section{Reduced-order models for the beam case}\label{app:romresdet}

This appendix gives the detailed equations governing the reduced-order dynamics in the beam case, in order to better understand which terms are present in the different ROMs, and which one are added when taking internal resonances into account. The discussion is restricted to the case of the the first NNM, where the main interesting feature is the presence of a 5:1 internal resonance (involving the nonlinear frequencies) between mode 1 and mode 3. Consequently ROMS have been built including these two modes as master coordinates. In the remainder, $R_1$ and $R_3$ will thus refer to the normal coordinates linked respectively to mode~1 and mode~3.

The simplest case is that of the ROM composed of a single master coordinate  $R_1$. The reduced dynamics reads, following Eq.~\eqref{eq:dynsingledof}:
\begin{equation}
\ddot{R}_1 + \omega_1^2 R_1 + (A_{111}^1 + h_{111}^1) R_1^3 + B_{111}^1 R_1 \dot{R}_1^2 \; = \; 0 \; . 
\label{eq:dynsingledofR1}
\end{equation}

The second case discussed in section~\ref{sec:beam} is that of a third-order normal form composed of modes 1 and 3, without taking into account any other terms due to the presence of an internal resonance. In this case the reduced dynamics deduces from Eq.~\eqref{eq:ROM} and directly reads:
\begin{subequations}\label{eq:dynROMo3R1R3}
\begin{align}
\ddot{R}_1 + \omega_1^2 R_1 & + (A_{111}^1 + h_{111}^1) R_1^3 + B_{111}^1 R_1 \dot{R}_1^2 \nonumber \\
& + 
\left(  A_{331}^1 + A_{313}^1 +  A_{133}^1 +  3h_{133}^1  \right) R_1 R_3^2 + 
B_{133}^1 R_1 \dot{R}_3^2  +
\left( B_{331}^1 + B_{313}^1  \right)\dot{R}_1 R_3 \dot{R}_3 = 0, 
\\
\ddot{R}_3 + \omega_3^2 R_3 & + (A_{333}^3 + h_{333}^3) R_3^3 + B_{333}^3 R_3 \dot{R}_3^2  \nonumber \\
& + 
\left(  A_{113}^3 + A_{131}^3 +  A_{311}^3 +  3h_{311}^3  \right) R_3 R_1^2 + B_{311}^3 R_3 \dot{R}_1^2  +      
\left( B_{113}^3 + B_{131}^3  \right)\dot{R}_3 R_1 \dot{R}_1 = 0.
\end{align}
\end{subequations}
These equations does not contain any invariant-breaking terms. Consequently they are unable to create a coupling between the invariant manifolds dynamics of mode 1 and 3. Since the ROM wants to follow the backbone curve of mode 1, adding the second equation on mode 3 without any invariant-breaking term is in this case meaningless, and this ROM will produce exactly the same backbone as Eq.~\eqref{eq:dynsingledofR1}. In order to counteract this property, the resonant monomials corresponding to the 1:3 internal resonance can be added to these equations, leading to the ROM with two modes and inclusion of new resonant terms. The reduced dynamics of this ROM now reads:
\begin{subequations}\label{eq:dynROMo3R1R3res13}
\begin{align}
\ddot{R}_1 + \omega_1^2 R_1 & + (A_{111}^1 + h_{111}^1) R_1^3 + B_{111}^1 R_1 \dot{R}_1^2 \nonumber \\
& + \left(A_{331}^1 + A_{313}^1 +  A_{133}^1 +  3h_{133}^1  \right) R_1 R_3^2 + 
B_{133}^1 R_1 \dot{R}_3^2  +      
\left( B_{331}^1 + B_{313}^1  \right)\dot{R}_1 R_3 \dot{R}_3 \nonumber \\
& + 
\left(A_{311}^1 + A_{113}^1 +  A_{131}^1 +  3h^1_{113}  \right)
R_1^2 R_3 +      
B_{311}^1 R_3 \dot{R}_1^2 + 
\left( B_{113}^1 +B_{131}^1 \right) R_1 \dot{R}_1 \dot{R}_3  
= 0, \label{eq:dynROMo3R1R3res13a}\\
\ddot{R}_3 + \omega_3^2 R_3 & + (A_{333}^3 + h_{333}^3) R_3^3 + B_{333}^3 R_3 \dot{R}_3^2  \nonumber \\
& + \left(  A_{113}^3 + A_{131}^3 +  A_{311}^3 +  3h_{311}^3  \right) R_1^2R_3  + B_{311}^3 R_3 \dot{R}_1^2  +      \left( B_{113}^3 + B_{131}^3  \right) R_1 \dot{R}_1 \dot{R}_3\nonumber \\
         &  + 
         (A_{111}^3 + h_{111}^3) R_1^3 + B_{111}^3 R_1 \dot{R}_1^2 = 0.\label{eq:dynROMo3R1R3res13b}
\end{align}
\end{subequations}
As explained in the main text, the only extra terms added to this ROM are the third line in each oscillator equation. For Eq.~\eqref{eq:dynROMo3R1R3res13a}, the cubic resonant monomial corresponding to 1:3 resonance is  $R_1^2 R_3$. This term needs to be accompanied with all his ancillary ones conveying the same resonance but with the velocity terms $\dot{R}_1$ and $\dot{R}_3$, resulting in the addition of the other resonant monomials $R_3 \dot{R}_1^2$ and $R_1 \dot{R}_1 \dot{R}_3$. For the second equation \eqref{eq:dynROMo3R1R3res13b}, the resonant monomial implying displacements only is $R_1^3$, which must also be complemented with all its companion terms sharing the same frequency resoannce and involcing velocities, thus the $R_1 \dot{R}_1^2$ term. One can note that $R_1^3$ on the second equation is an invariant-breaking term. Consequently this system is prone to activate the resonance and break the invariance property of the single-mode manifold.

The last system investigated includes as extra terms those corresponding to the 1:1 resonance. This final two-dofs reduced dynamics finally reads:
\begin{subequations}\label{eq:dynROMo3R1R3res1311}
\begin{align}
\ddot{R}_1 + \omega_1^2 R_1 & + 
(A_{111}^1 + h_{111}^1) R_1^3 + B_{111}^1 R_1 \dot{R}_1^2 
\nonumber \\
& + 
\left(A_{331}^1 + A_{313}^1 +  A_{133}^1 +  3h_{133}^1  \right) R_1 R_3^2 + 
B_{133}^1 R_1 \dot{R}_3^2  +      
\left( B_{331}^1 + B_{313}^1  \right)\dot{R}_1 R_3 \dot{R}_3 \nonumber \\
& + 
\left(A_{311}^1 + A_{113}^1 +  A_{131}^1 +  3h^1_{113}  \right)
R_1^2 R_3 +      
B_{311}^1 R_3 \dot{R}_1^2 + 
\left( B_{113}^1 +B_{131}^1 \right) R_1 \dot{R}_1 \dot{R}_3  
\nonumber \\
& +
+(A_{333}^1 + h_{333}^1) R_3^3 + B_{333}^1 R_3 \dot{R}_3^2 
= 0, \\
\ddot{R}_3 + \omega_3^2 R_3 & + 
(A_{333}^3 + h_{333}^3) R_3^3 + B_{333}^3 R_3 \dot{R}_3^2  
\nonumber \\
& + 
\left(  A_{113}^3 + A_{131}^3 +  A_{311}^3 +  3h_{311}^3  \right) R_1^2R_3  + 
B_{311}^3 R_3 \dot{R}_1^2  + 
\left( B_{113}^3 + B_{131}^3  \right) R_1 \dot{R}_1 \dot{R}_3
\nonumber \\
& + 
\left(A_{331}^3 + A_{313}^3 +  A_{133}^3 +  3h_{133}^3  \right) R_1 R_3^2 + 
B_{133}^3 R_1 \dot{R}_3^2  +      
\left( B_{331}^3 + B_{313}^3  \right)\dot{R}_1 R_3 \dot{R}_3 \nonumber \\
&  + 
(A_{111}^3 + h_{111}^3) R_1^3 + B_{111}^3 R_1 \dot{R}_1^2 = 0.
\end{align}
\end{subequations}
This system now contains all the possible cubic monomials. Consequently it is fully equivalent to the system obtained by applying the second-order normal form only.

\end{document}